\documentclass[a4paper,fleqn,usenatbib]{mnras}

\usepackage{newtxtext,newtxmath}

\usepackage[T1]{fontenc}
\usepackage{ae,aecompl}

\usepackage{graphicx}	
\usepackage{amsmath}	
\usepackage{amssymb}	
\usepackage{amsfonts}
\usepackage{epstopdf}
\usepackage{pdfpages}
\usepackage{epsfig}
\DeclareGraphicsExtensions{.pdf,.png,.jpg}
\usepackage{tabularx}
\usepackage{supertabular}
\usepackage{longtable}
\usepackage{rotating}
\usepackage{bold-extra}
\usepackage{appendix}
\usepackage{listings}
\usepackage{float}
\usepackage{natbib}
\def\apjl{ApJL}
\def\aj{AJ}
\def\apj{ApJ}
\def\mnras{MNRAS}
\def\apjs{ApJS}
\def\araa{ARAA}
\def\aap{A\&A}

\def\aaps{A\&AS}
\def\nat{Nature}

\usepackage{url}
\usepackage{hyperref}
\usepackage{enumerate}
\newcommand{\comt}[1]{\ignorespaces}
\usepackage{morefloats}
\usepackage{pdflscape}
\usepackage{afterpage}
\usepackage{lipsum}

\def\Msun{M$_{\odot}$}

\def\submm{submillimetre}
\def\uJy{$\mu$Jy}
\def\overdens{$2.8\pm0.5$}

\title[A SCUBA-2 survey of $z\sim0.8$--$1.6$ clusters]{The submillimetre view of massive clusters at $z\sim0.8$--$1.6$}

\author[Cooke et al.] {E. A. Cooke$^{1}$\thanks{elizabeth.a.cooke@durham.ac.uk},
Ian Smail$^{1,2}$,
S.\ M.\ Stach$^{1}$,
A.\ M.\ Swinbank$^{1,2}$,
R.\ G.\ Bower$^{1,2}$,
\newauthor Chian-Chou Chen$^3$, 
Y.\ Koyama$^4$,
A.\ P.\ Thomson$^5$
\\
$^1$Centre for Extragalactic Astronomy, Department of Physics, Durham University, Durham, DH1 3LE, UK \\
$^2$Institute for Computational Cosmology, Department of Physics, University of Durham, South Road, Durham, DH1 3LE, UK\\
$^3$European Southern Observatory, Karl Schwarzschild Strasse 2, Garching, Germany\\
$^4$Subaru Telescope, National Astronomical Observatory of Japan, National Institutes of Natural Sciences, 650 North A'ohoku Place, Hilo, HI 96720, USA\\
$^5$Jodrell Bank Centre for Astrophysics, School of Physics and Astronomy, The University of Manchester, Oxford Road, Manchester, M13 9PL, UK\\
}

\date{Accepted XXX. Received YYY; in original form ZZZ}

\pubyear{2019}

\begin{document}
\label{firstpage}
\pagerange{\pageref{firstpage}--\pageref{lastpage}}
\maketitle

\begin{abstract}
We analyse $850$\,\micron\ continuum observations of eight massive X-ray detected galaxy clusters at $z\sim0.8$--$1.6$ taken with SCUBA-2 on the James Clerk Maxwell Telescope. We find an average overdensity of $850$\,\micron-selected sources of a factor of $4\pm2$ per cluster within the central $1$\,Mpc compared to the field. We investigate the multiwavelength properties of these sources and identify $34$ infrared counterparts to $26$ SCUBA-2 sources. Their colours suggest that the majority of these counterparts are probable cluster members. We use the multi-wavelength far-infrared photometry to measure the total luminosities and total cluster star-formation rates demonstrating that they are roughly three orders of magnitude higher than local clusters. We predict the $H$-band luminosities of the descendants of our cluster \submm\ galaxies and find that their stellar luminosity distribution is consistent with that of passive elliptical galaxies in $z\sim0$ clusters. Together, the faded descendants of the passive cluster population already in place at $z\sim1$ and the cluster \submm\ galaxies are able to account for the total luminosity function of early-type cluster galaxies at $z\sim0$. This suggests that the majority of the luminous passive population in $z\sim0$ clusters are likely to have formed at $z\gg1$ through an extreme, dust-obscured starburst event. 

\end{abstract}


\section{Introduction} \label{sec:intro}

In the local Universe, the most massive galaxies reside in the centres of galaxy clusters \citep[e.g.,][]{Kauffmann2004,Bamford2009}. These massive galaxies typically have little-to-no ongoing star formation, display spheroidal ``early-type" morphologies and contain old, metal-rich stellar populations. Detailed ``archaeological" studies of the star-formation histories of luminous ellipticals ($\gtrsim L^*$) indicate most of their stars were formed $8$--$11$\,Gyr ago at $z>1$, through a series of bursts of star formation \citep[e.g.,][see also \citealt{Johnston2014,Cooke2015}]{Bower1992,Thomas2005,Citro2016,GonzalezDelgado2017}.  More massive galaxies have been found to have older stellar populations \citep[e.g.,][]{Nelan2005}, and ellipticals within clusters have older stellar ages than those residing in the field \citep[e.g.,][]{Rettura2011}. 

At $z\sim1$ the cores of massive clusters appear to have already formed, and display similar properties to local clusters \citep[e.g.,][]{Cerulo2016}. Indeed, there are now several known examples of apparently passive cluster cores at $z\sim1$--$2$ \citep[e.g.,][]{Strazzullo2013,Newman2014,Cooke2016,Lee-Brown2017}. However, there is also evidence that significant star formation is occurring within clusters at these epochs \citep[e.g.,][]{Hayashi2010, Tran2010, Zeimann2013, Brodwin2013}. In particular, the number of dusty star-forming galaxies in clusters, observable by their bright infrared luminosities, increases out to $z > 1$ \citep[e.g.,][]{Best2002,Webb2005,Geach2006,Tran2010,Popesso2012,Webb2013,Alberts2014,Noble2016}. This dust-obscured star formation traces increased activity in the clusters, whose mean integrated star-formation rate (SFR) appears to evolve very rapidly: $\sim(1+z)^{\gamma}$, with $\gamma\sim6$--$7$ \citep{Kodama2004,Geach2006,Koyama2010,Koyama2011,Shimakawa2014,Ma2015}, compared to the field: $\gamma=4$ \citep{Ilbert2015}. This accelerated star formation activity means that by $z\sim1$--$1.5$, clusters host significant numbers (although with large cluster-to-cluster variation) of dusty ``\submm\ galaxies" (SMGs), so-called because of their bright luminosities at \submm\ wavelengths, \citep[e.g.,][]{Tadaki2012,Ma2015}. 

\begin{figure*} 
\includegraphics[trim=2cm 0.5cm 0cm 0.5cm,clip,width=0.27\textwidth]{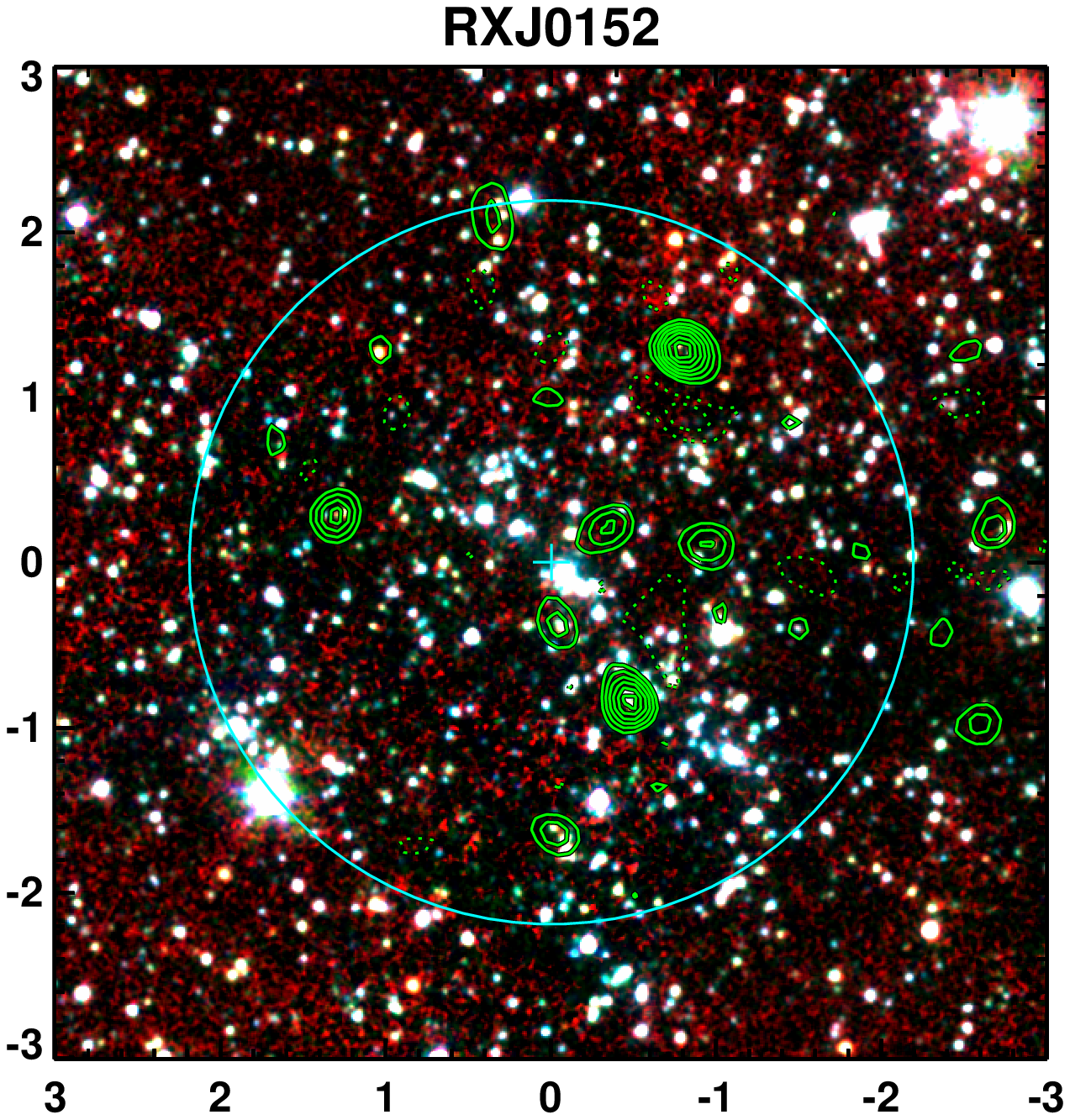}
\includegraphics[trim=2cm 0.5cm 0cm 0.5cm,clip,width=0.27\textwidth]{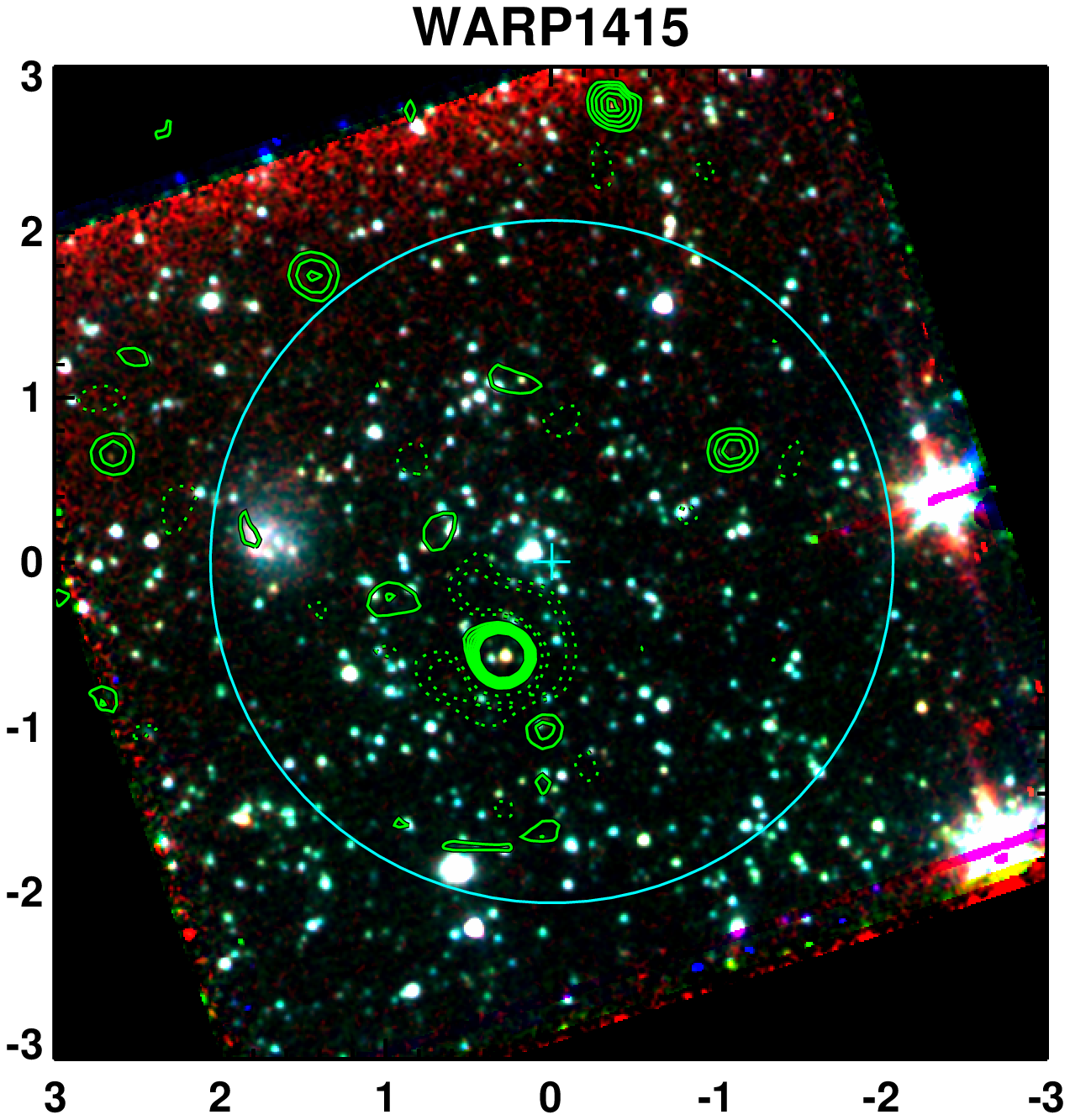}
\includegraphics[trim=2cm 0.5cm 0cm 0.5cm,clip,width=0.27\textwidth]{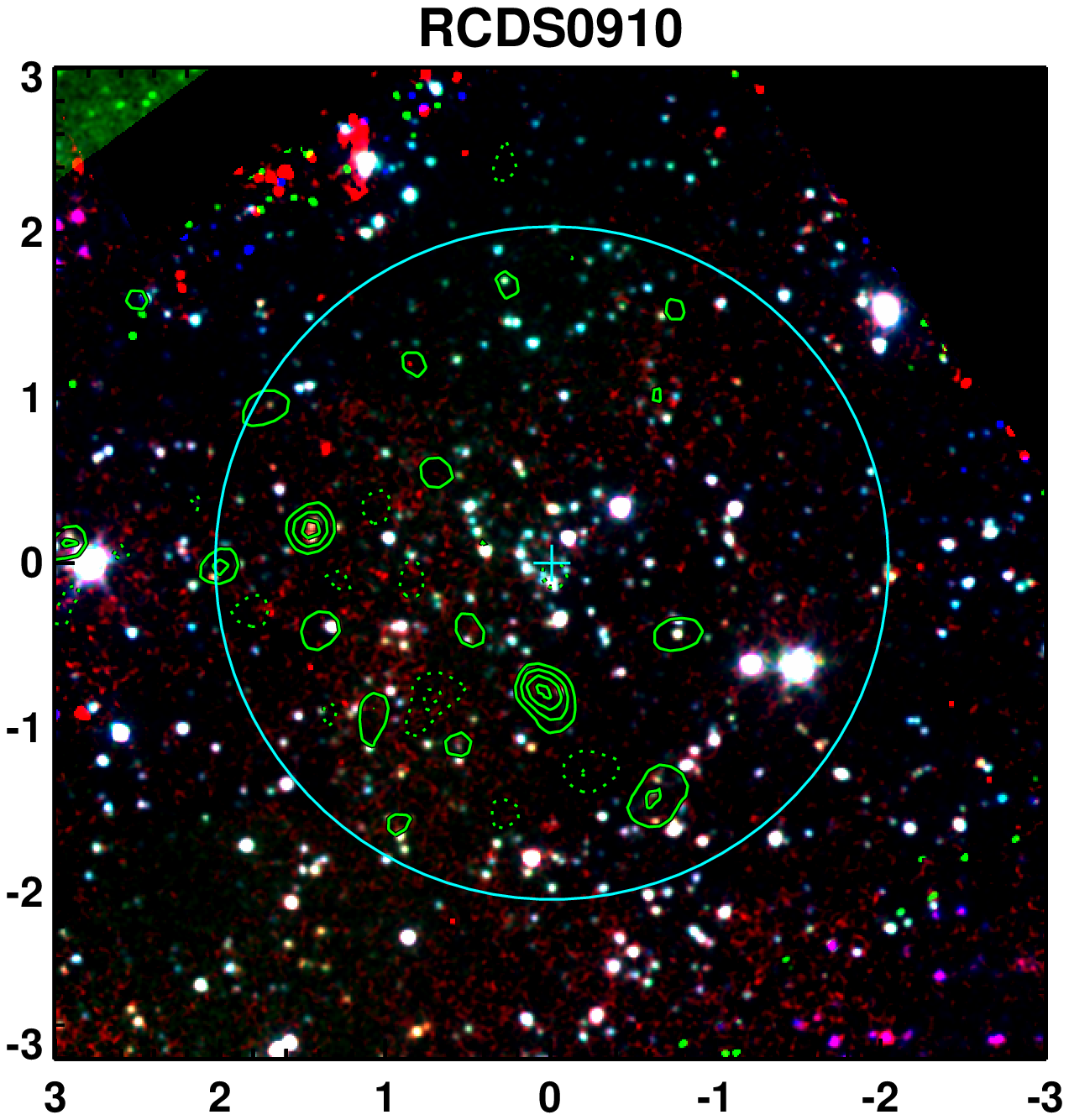}
\includegraphics[trim=2cm 0.5cm 0cm 0.5cm,clip,width=0.27\textwidth]{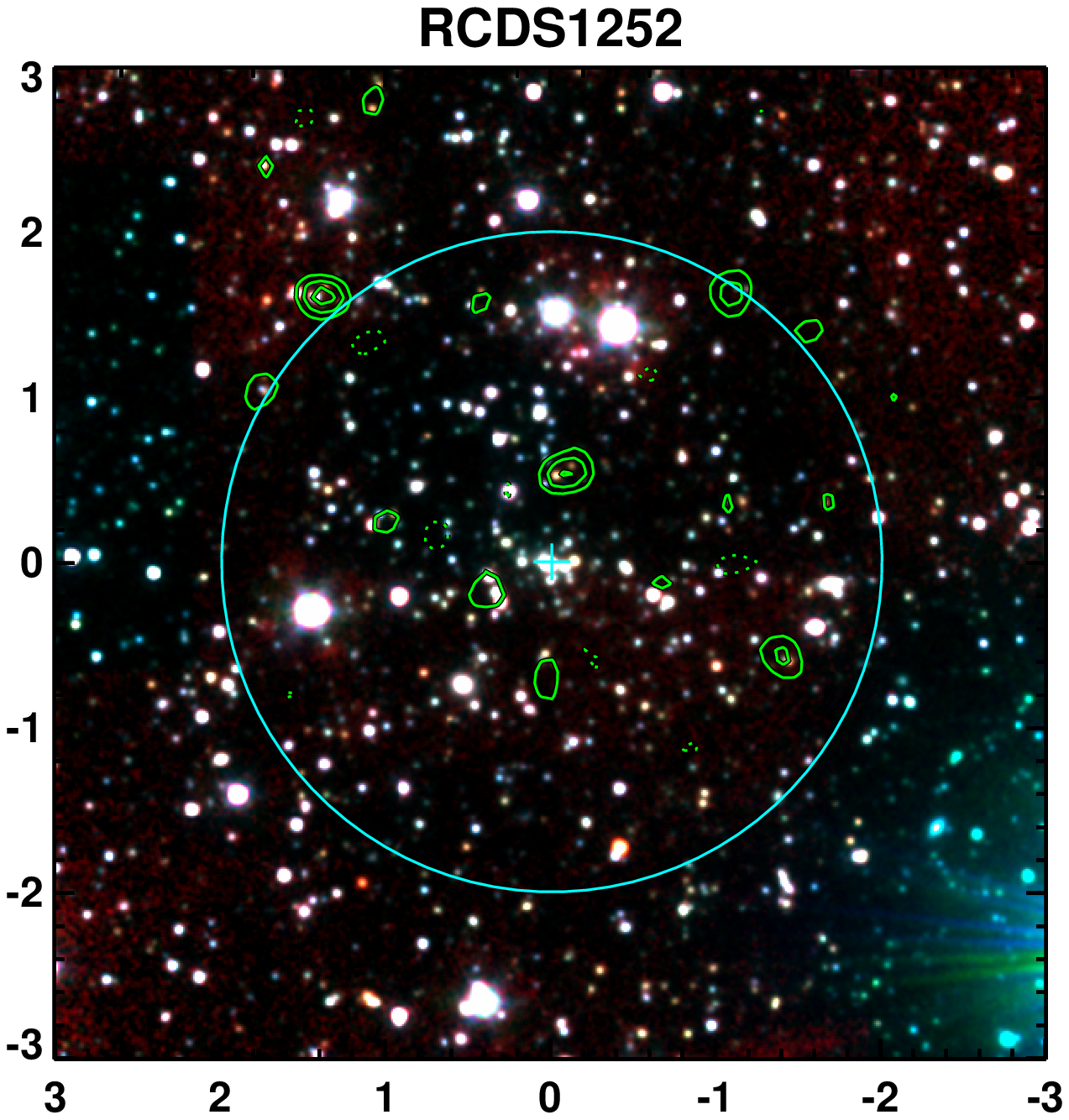}
\includegraphics[trim=2cm 0.5cm 0cm 0.5cm,clip,width=0.27\textwidth]{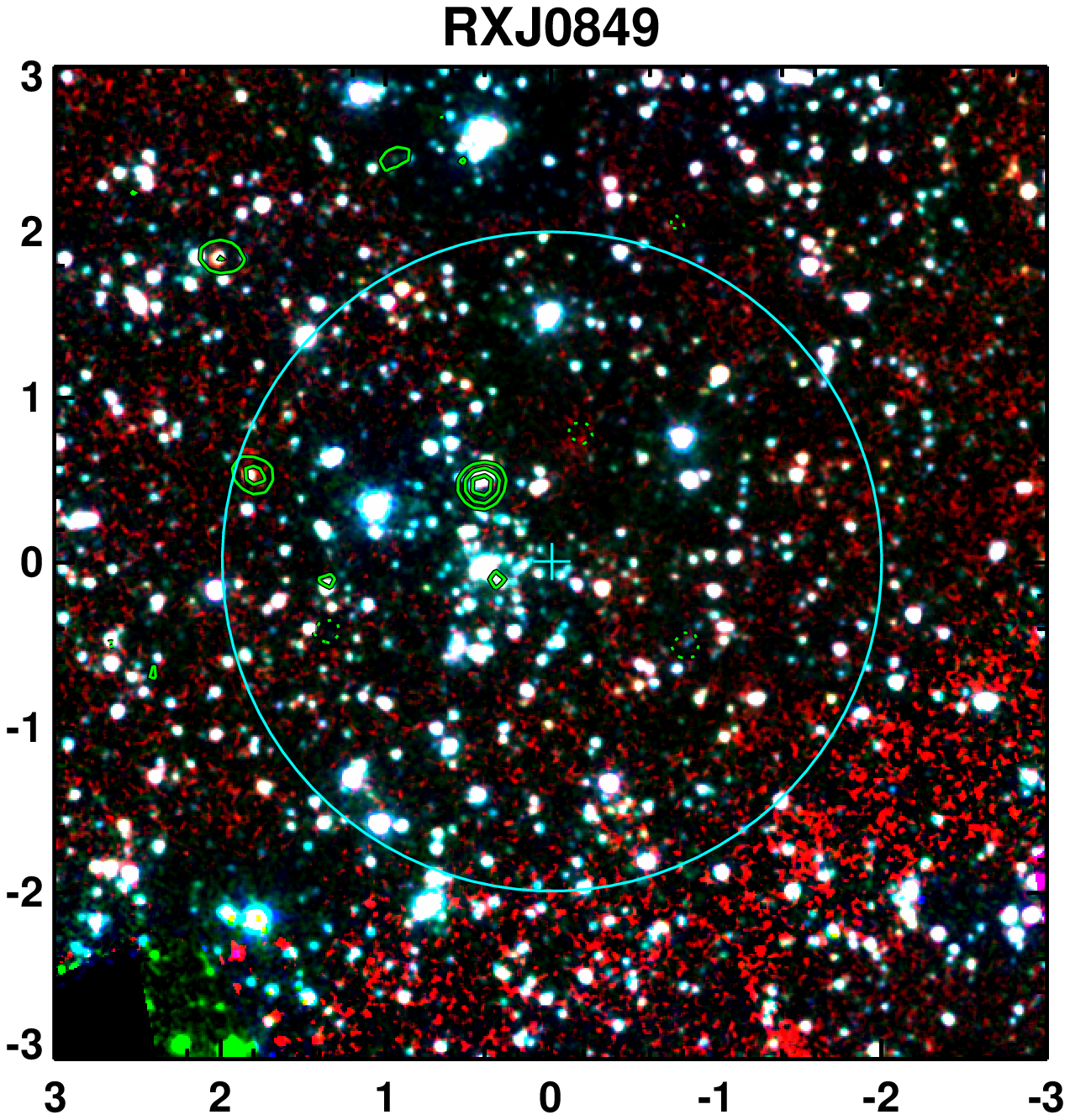}
\includegraphics[trim=2cm 0.5cm 0cm 0.5cm,clip,width=0.27\textwidth]{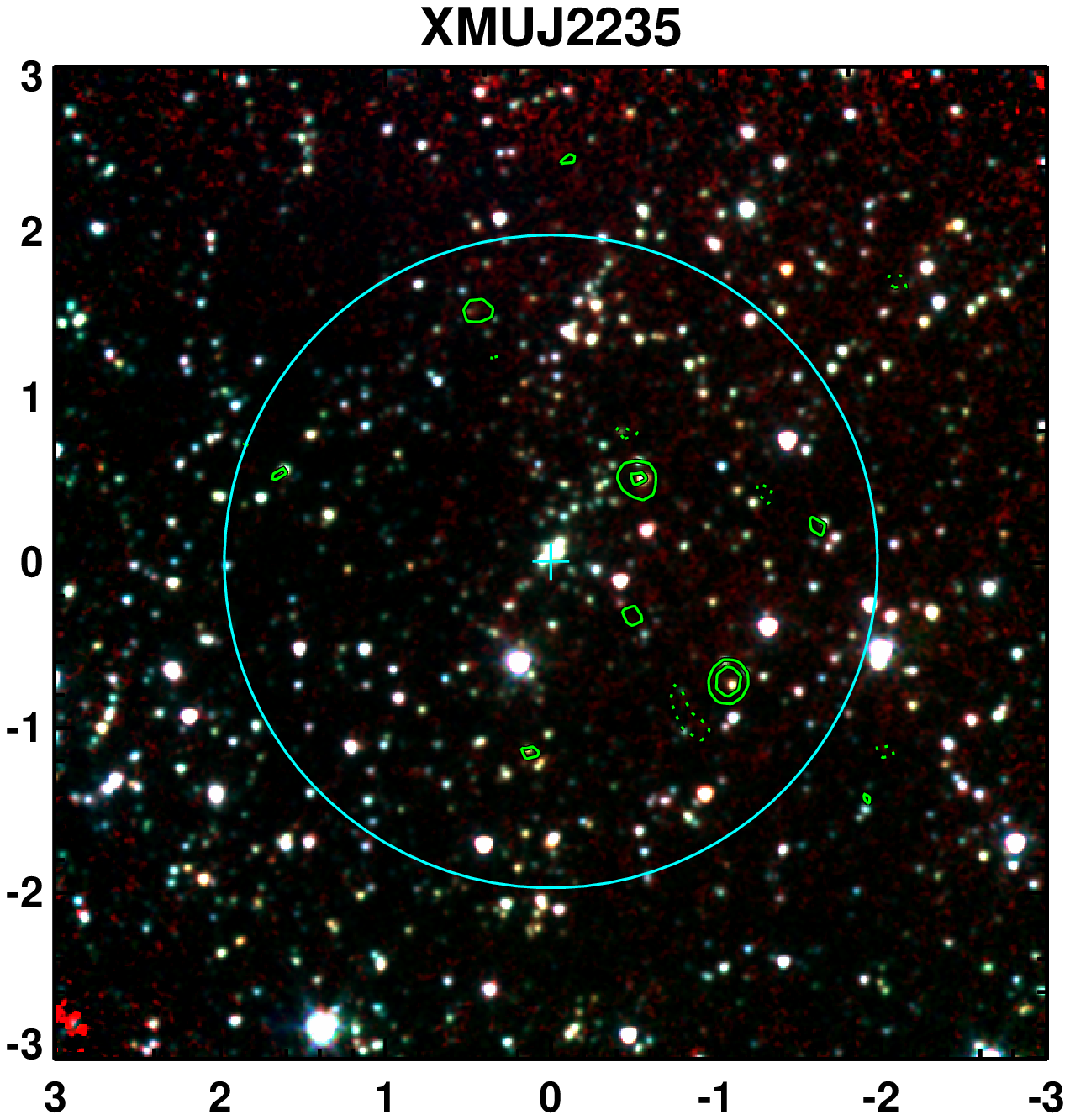}
\includegraphics[trim=2cm 0.5cm 0cm 0.5cm,clip,width=0.27\textwidth]{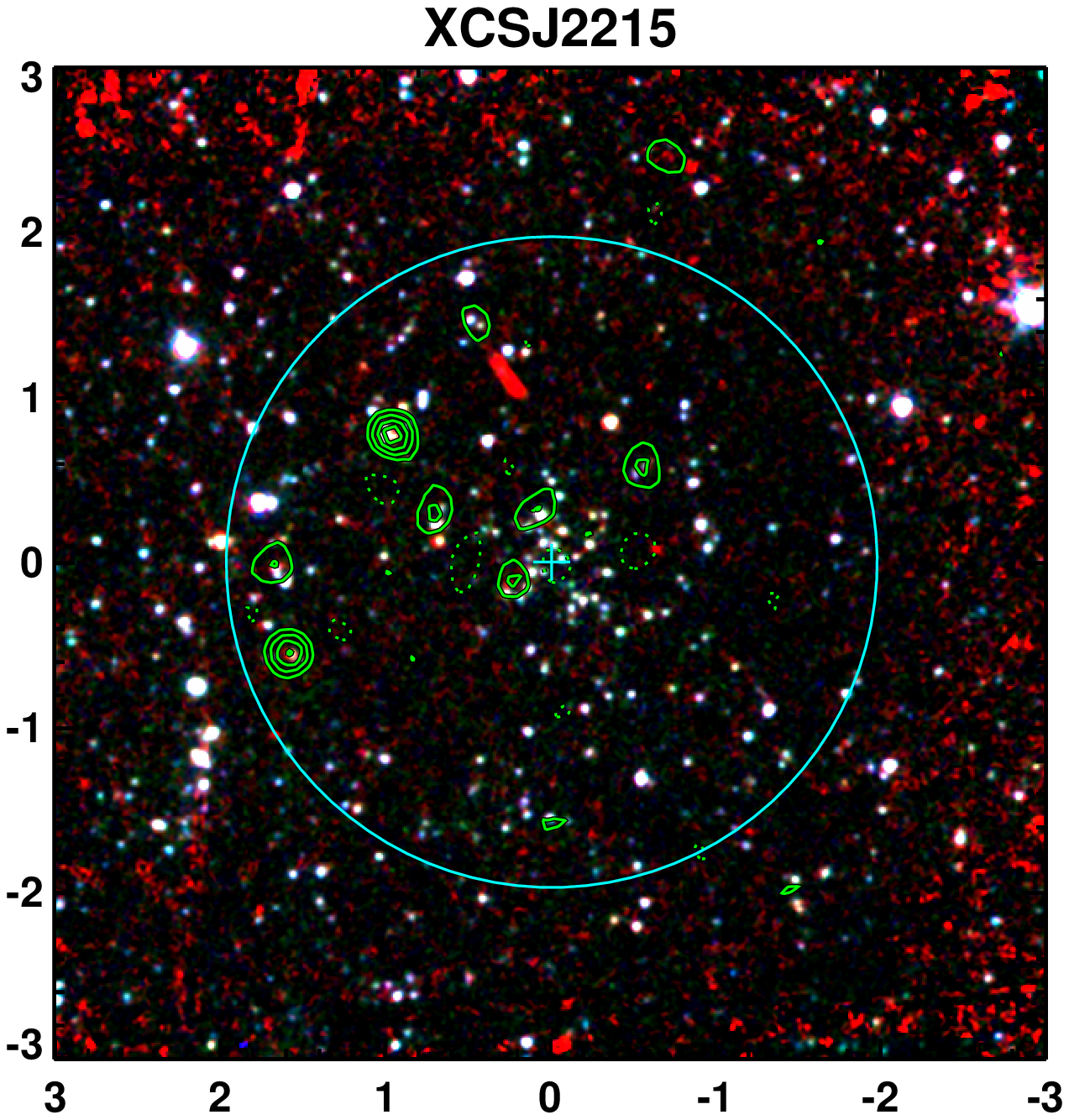}
\includegraphics[trim=2cm 0.5cm 0cm 0.5cm,clip,width=0.27\textwidth]{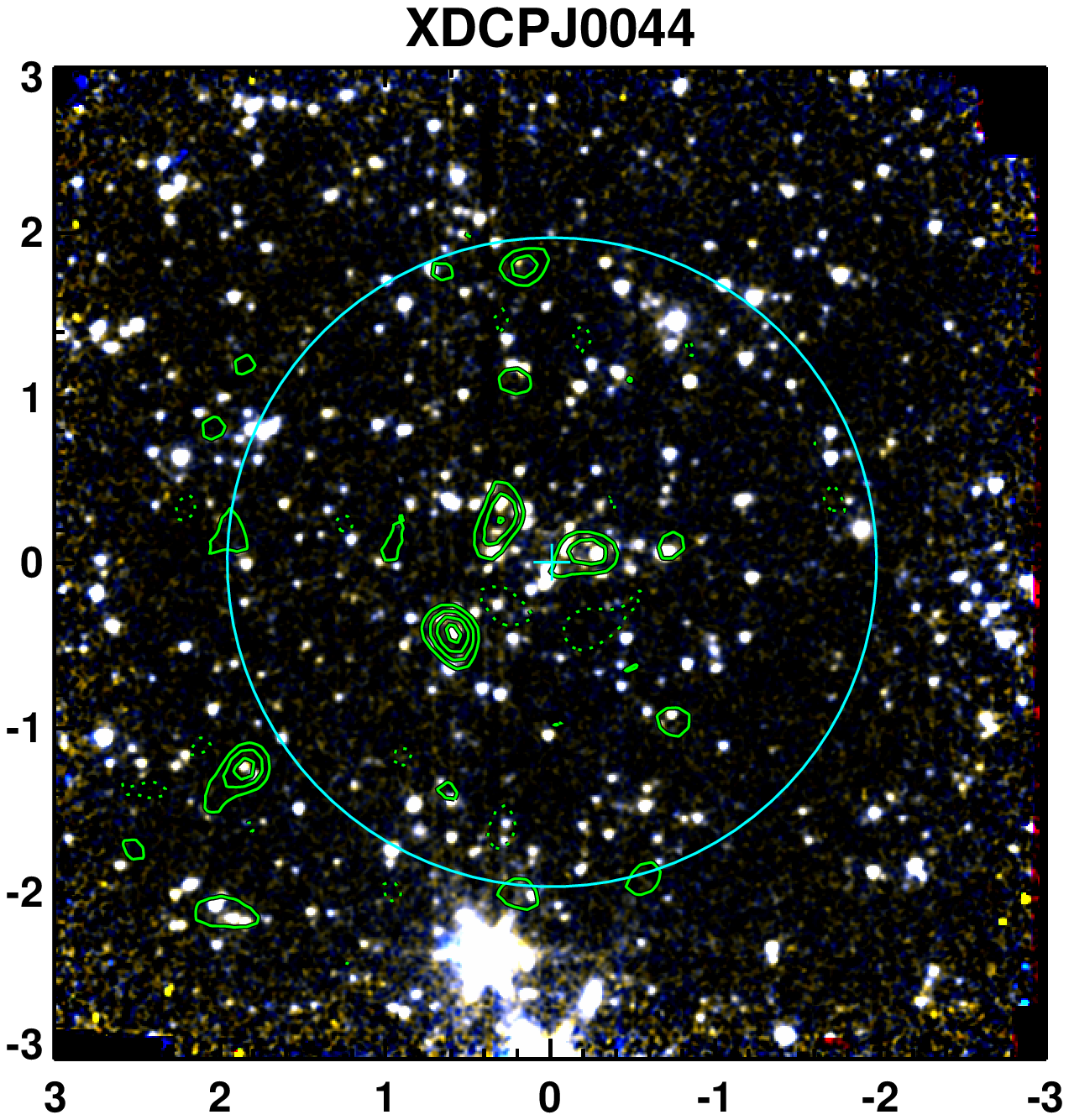}
\caption{IRAC 3.6\,\micron, 4.5\,\micron, and 5.8\,\micron\ three-colour images of the eight clusters in our sample. XDCP\,J0044 only has coverage in two IRAC channels so it is a false-colour 3.6\,\micron\ and 4.5\,\micron\ image. The contours show the SCUBA-2 $850$\,\micron\ flux; solid contours start at $3\sigma$ and increase by $2\sigma$. The dotted contours show $-3\sigma$ and $-5\sigma$. The X-ray detected centre of each cluster is marked by a cross. The large circle denotes $1$\,Mpc radius around each cluster core. We detect $83$ \submm\ sources with $S/N \geq 4$ or $S_{850}\geq4$\,mJy. }
\label{fig:colourimgs}
\end{figure*}

\begin{figure*} 
\begin{centering}
\includegraphics[trim=0cm 0.2cm 1.6cm 1.2cm,clip,width=\textwidth]{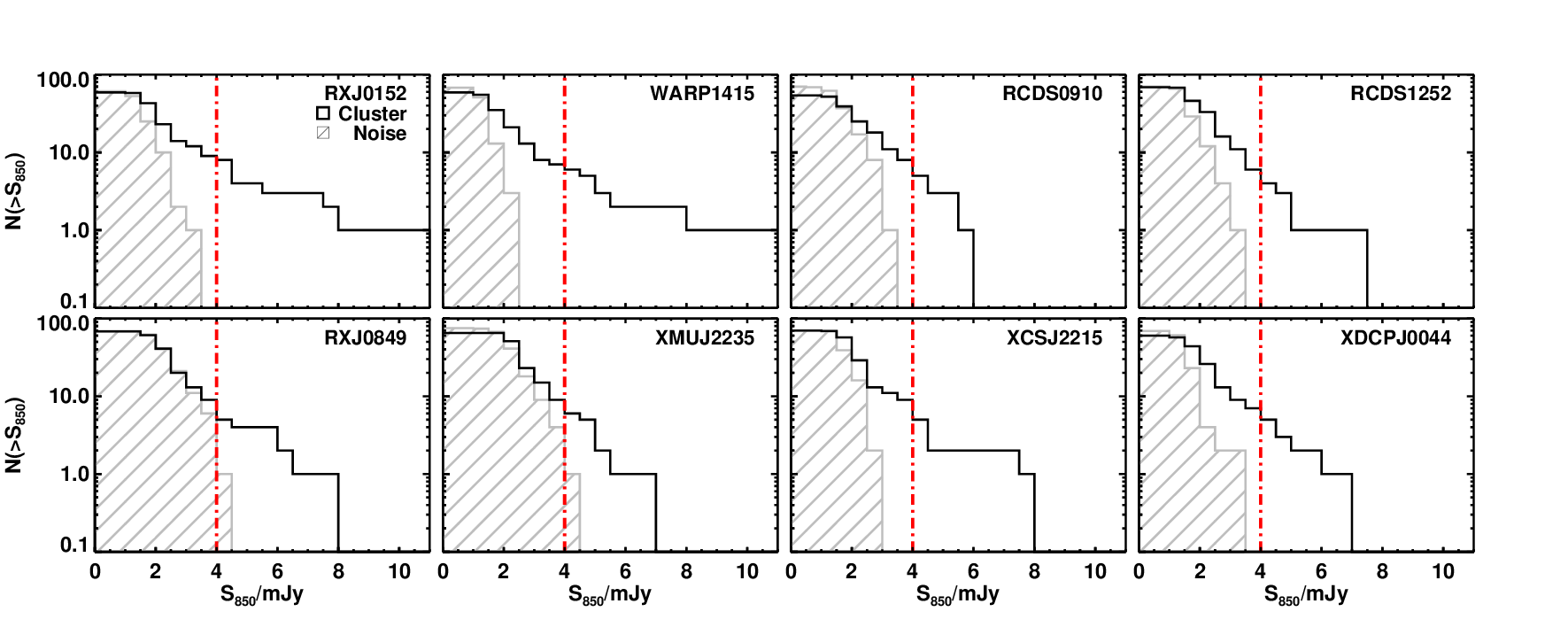}
\caption{Cumulative number counts of \submm\ sources above a given flux density for all eight clusters. Sources detected in the SCUBA-2 maps are shown in black and sources in the jack-knifed maps are shown by the hatched grey histogram. The dot-dashed line shows the uniform $4$\,mJy cut we use to select sources, corresponding to $>$$3\sigma$ across all the maps. We also include all sources with a signal-to-noise ratio $S/N\geq4$, which gives an overall false-detection rate of $\sim$$13\%$.  }
\label{fig:fpr}
\end{centering}
\end{figure*}

Studies of the dusty star-forming population of cluster galaxies at $z>1$ has been hampered by limited statistics, which often show starkly different results. For example, \citet{Smail2014} showed that the core of Cl\,J0218$-$0510 at $z =1.62$ \citep{Papovich2010,Tanaka2010} is mostly inactive. Using the SCUBA-2 Cosmology Legacy Survey (S2CLS) map of the UKIDSS Ultra Deep Survey \citep[UDS;][]{Geach2017} they identified $31$ probable cluster SMGs, but found that few of these lie in the core ($2$\,Mpc diameter region), which is instead dominated by apparently passive, massive red galaxies \citep[$>L^*$ or $M_H\sim-23$;][]{Lotz2013,Hatch2016,Hatch2017,Lee-Brown2017}. By contrast, \citet{Stach2017} and \citet{Hayashi2017} revealed $14$ luminous far-infrared galaxies, with a combined star-formation rate of $>$$1000$\,\Msun\,yr$^{-1}$, within a $500$\,kpc diameter region in the heart of XCS\,J2215.9$-$1738 at $z=1.46$ \citep{Stanford2006,Hilton2010}. These observations mean that XCS\,J2215 is one of the most strongly star-forming clusters known to date \citep{Stach2017,Hayashi2018}. These two clusters illustrate the wide variation in the dusty star-forming population seen in distant clusters and the environments in which these systems are found. 

Untangling the average evolution of massive cluster galaxies, and their likely form at high redshift, requires a larger, more uniform sample of clusters. To this end, we have undertaken a sensitive $450$/$850$\,\micron\ survey of eight massive galaxy clusters at $0.8<z<1.6$ with SCUBA-2 \citep{Holland2013} on the James Clerk Maxwell Telescope (JCMT) to search for dust-obscured highly star-forming galaxies within these massive structures and statistically measure the properties of cluster SMGs. 

The layout of this paper is as follows: Section \ref{sec:data} outlines our observations and data reduction. In Section \ref{sec:properties} we analyse the mid- to far-infrared properties of the observed \submm\ sources. Section \ref{sec:discussion} discusses the star-formation rates and evolution of cluster SMGs with redshift. Our conclusions are presented in Section \ref{sec:concl}. Throughout we use a $\Lambda$CDM cosmology with $H_0=70$\,km\,s$^{-1}$\,Mpc$^{-1}$, $\Omega_M=0.3$ and $\Omega_\Lambda=0.7$. Magnitudes are given in the AB system.

\section{Observations and Data Reduction} \label{sec:data}
Our sample is comprised of eight well-studied $z\gtrsim1$ X-ray detected clusters. The X-ray detections are required to ensure the selection of bound systems, and enable estimates of cluster mass and dynamical structure. The eight clusters (Table \ref{table:clusters}) were observed with SCUBA-2 simultaneously at $450$\,\micron\ and $850$\,\micron\ in weather conditions $\tau_{250{\rm GHz}} < 0.08$ between 2013 April 09 and 2016 May 10. In total each cluster was observed for an average of $10$\,hours using a standard constant-velocity daisy mapping pattern. The sensitivity in the resulting maps drops to 50\% at a radius of $\sim$$5.4$\,arcmin from the map centre due to the scan coverage of the daisy pattern.

Individual maps for each night of observation were reduced using the Dynamic Interactive Map-Maker ({\sc dimm}) tool of the Sub-Millimetre User Reduction Facility \citep[{\sc smurf};][]{smurf} with the blank field configuration in order to detect point sources within the maps. The maps were calibrated using a flux conversion factor of ${\rm FCF}_{450\micron} =  491$\,Jy\,beam$^{-1}$\,pW$^{-1}$ and ${\rm FCF}_{850\micron} =  537$\,Jy\,beam$^{-1}$\,pW$^{-1}$ and then combined using inverse-variance weighting to create a final map per cluster at each wavelength. To improve point source detection, the resulting $450$\,\micron\ and $850$\,\micron\ maps were match-filtered with an $8$\,arcsec and $15$\,arcsec Gaussian filter, respectively. This match-filtering step in the data reduction has been shown to introduce a small ($10\%$) loss of flux from point sources \citep[e.g.,][]{Chen2013,Geach2017}. We thus apply an upward correction of $10\%$ to our measured fluxes. 

At $850$\,\micron\ the median noise in the centre of the maps is $0.6^{+0.3}_{-0.1}$\,mJy (Table \ref{table:clusters}). The maps were cropped to radii of $2.5$\,arcmin, where the noise properties of the maps are low and more uniform (with a variation across the map of $\sim33\%$). This radius corresponds to approximately $1.2$\,Mpc at $z\sim1.2$, the median redshift of our sample. False-colour images of the eight clusters are shown in Figure \ref{fig:colourimgs}. These clusters show a wide range of activity at $850$\,\micron\ within the central $1$\,Mpc. 

\begin{table*}
\caption{Properties of the eight clusters in our sample. The final two columns indicate the number of \submm\ sources selected in each field (Table \ref{table:S2}), and the number of infrared-selected counterparts, including those with multiple counterparts (Table \ref{table:IR}). }
\begin{tabular}{ lcccccccc}
\hline
\multicolumn{1}{c}{ID} &
\multicolumn{1}{c}{R.A.} &
\multicolumn{1}{c}{Decl.} &
\multicolumn{1}{c}{$z$} &
\multicolumn{1}{c}{$\sigma_{850}$} &
\multicolumn{1}{c}{kT$_{\rm X}$} &
\multicolumn{1}{c}{$\log_{10}(M_{200})$} &
\multicolumn{1}{c}{$N_{850}$} &
\multicolumn{1}{c}{$N_{\rm IR}$} \\
\multicolumn{1}{c}{} &
\multicolumn{2}{c}{[J2000]} &
\multicolumn{1}{c}{} &
\multicolumn{1}{c}{[mJy\,beam$^{-1}$]} &
\multicolumn{1}{c}{[eV]} &
\multicolumn{1}{c}{[$\log_{10}({\rm M}_{\odot})$]} &
\multicolumn{1}{c}{} &
\multicolumn{1}{c}{} \\
\hline
\hline
RXJ0152$-$1357 & 01:52:44.18 & $-$13:57:15.8 & 0.831 & 0.51 & 4.3$\pm$0.5 & 14.5$\pm$0.1  & 14 &  8 \\
WARPJ1415$+$3611 & 14:15:10.48 & $+$36:11:59.0 & 1.030 & 0.49 & 6.2$\pm$0.8 & 14.7$\pm$0.1  & 12 &  3 \\
RDCSJ0910$+$5422 & 09:10:44.90 & $+$54:22:08.9 & 1.100 & 0.59 & 7.0$\pm$2.0 & 14.8$\pm$0.2  & 14 &  5 \\
RDCSJ1252$-$2927 & 12:52:54.40 & $-$29:27:17.0 & 1.237 & 0.63 & 6.0$\pm$0.7 & 14.6$\pm$0.1  & 10 &  2 \\
RXJ0849$+$4451 & 08:48:56.20 & $+$44:52:00.0 & 1.261 & 0.89 & 6.0$\pm$3.0 & 14.6$\pm$0.3  &  5 &  2 \\
XMUJ2235$-$2557 & 22:35:20.60 & $-$25:57:42.0 & 1.393 & 0.89 & 6.0$\pm$3.0 & 14.6$\pm$0.3  &  6 &  2 \\
XCSJ2215$-$1738 & 22:15:58.51 & $-$17:38:02.5 & 1.450 & 0.64 & 7.0$\pm$3.0 & 14.7$\pm$0.3  & 10 & 10 \\
XDCPJ0044$-$2033 & 00:44:05.20 & $-$20:33:59.7 & 1.579 & 0.54 & 7.0$\pm$1.0 & 14.6$\pm$0.2  & 12 &  2 \\
\hline
\end{tabular}
\label{table:clusters}
\end{table*}

\subsection{Source selection}
To select \submm\ sources from the SCUBA-2 maps we first use {\sc aegean} \citep{Hancock2012,Hancock2018} to identify peaks brighter than $1\sigma$ above the noise in each map and then measure fluxes from the SCUBA-2 maps. As these sources are unresolved we then take as the flux of each source its peak flux value in each of the $450$\,\micron\ and $850$\,\micron\ maps. The error on this flux is the value in the corresponding pixel in the error map produced from the data reduction pipeline. 
Owing to the better uniformity of our $850$\,\micron\ maps we concentrate on those in the following analysis and primarily use the $450$\,\micron\ data to constrain the spectral energy distributions of the $850$\,\micron\ sources. 

Jack-knifed maps were created using the same process detailed above but, before mosaicking, half of the individual scans were inverted in order to create maps with no astronomical signal \citep[e.g.,][]{Weiss2009,Geach2013}. We then run our detection process detailed above on these jack-knifed maps in order to estimate the contamination expected from false positive sources in each map. Figure \ref{fig:fpr} shows the distribution of sources as a function of their $850$\,\micron\ flux for each cluster and the jack-knifed maps. 

To select our sample, we apply a uniform cut of $S/N = S_{850}/\sigma_{850} \geq 4$, which selects $79$ \submm\ sources and corresponds to a false detection rate in the jack-knifed maps of $11\%$. 
To construct a flux-limited sample, we also include an ``extended sample", where we select all sources with $S_{850}\geq 4$\,mJy. This includes a further $4$ sources between $4\sigma$ and $3\sigma$, although with a false detection rate of $2$, i.e. $50\%$. 

Our final \submm\ sample has $83$ sources detected at either ${\rm S/N}\geq4\sigma$ or $S_{870} \geq 4$\,mJy across all cluster fields, with an overall expected false detection rate of $13\%$. The properties of the full sample are listed in Table \ref{table:S2}.

\subsection{Number counts}\label{sec:ncounts}
\begin{figure}
\begin{centering}
\includegraphics[trim=0.8cm 0.3cm 1.2cm 0cm,clip,width=0.85\columnwidth]{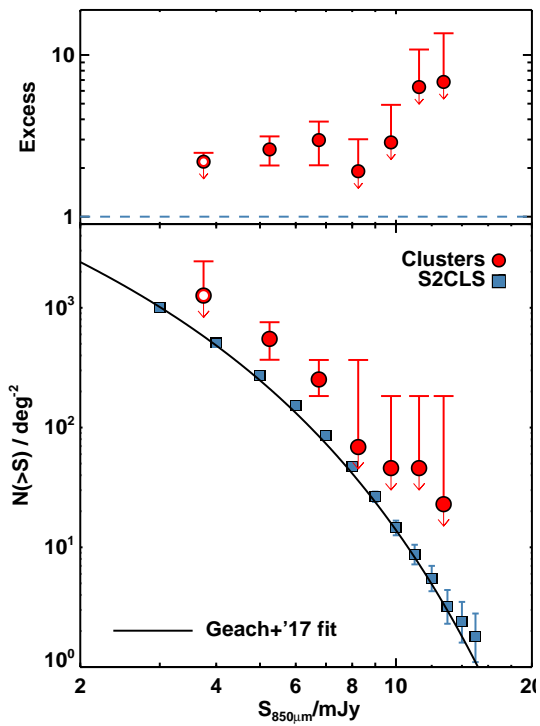}
\caption{\emph{Bottom:} Cumulative number counts for \submm\ sources detected in the central $2.5$\,arcmin ($1.2$\,Mpc) radius of the cluster fields with observed fluxes $S_{850}\geq 4$\,mJy. Fluxes have been deboosted following \citet{Geach2017} and corrected for incompleteness and the expected false-positive rate ($13\%$ at $S_{850}^{\rm boosted}\geq 4$\,mJy). Error bars show the $16^{\rm th}$ and $84^{\rm th}$ percentiles of number densities between individual cluster fields. Upper limits are given where the $16^{\rm th}$ percentile lower limit includes number densities of zero. The open symbol shows the number counts below our flux limit; there is information here due to the flux deboosting correction. For comparison we also show the number counts from the SCUBA-2 survey of the UDS field. \emph{Top:} Cluster counts divided by field counts showing the excess of \submm\ sources in the clusters. Error bars show Poisson errors on the number counts. These $z=0.8$--$1.6$ clusters show an overdensity of a factor of \overdens\ relative to the field within their central $2.4$\,Mpc (diameter). 
}
\label{fig:ncounts}
\end{centering}
\end{figure}

To calculate the expected completeness of our sample, we insert fake sources into the jack-knifed $850$\,\micron\ maps for each cluster field. Fake sources are randomly placed within the maps and have fluxes distributed according to:
\begin{equation}
\qquad \frac{dN}{dS} = \left(\frac{N_0}{S_0}\right) \left(\frac{S}{S_0}\right)^\gamma \exp\left(-\frac{S}{S_0}\right)
\end{equation}

\noindent with $N_0 = 7180$\,deg$^{-2}$, $S_0 = 2.5$\,mJy and $\gamma = 1.5$ \citep{Geach2017}. We then run our source detection as described above and include a source as recovered if a point source is found within the full-width at half-maximum (FWHM) of the SCUBA-2 $850$\,\micron\ effective beam ($14.6$\,arcsec). This is repeated $1000$ times per map, giving a sample of $8000$ fake sources. We then evaluate the recovery rate of fake sources as a function of $850$\,\micron\ flux and use this to correct our observed number counts, shown in Figure \ref{fig:ncounts}. 

We also apply a flux-deboosting correction appropriate for SCUBA-2 \citep[see][]{Geach2017}, which statistically corrects for the fact that an individual source's flux may be overestimated due to noise in the map\footnote{We have tested this deboosting correction on each cluster jack-knifed map with our catalogue of fake sources and find the power law derived in \citet{Geach2017} provides a good fit to our data. }. Figure \ref{fig:ncounts} shows the cumulative number counts of $850$\,\micron-selected sources in the cluster fields compared to field counts from the S2CLS/UDS \citep{Geach2017}. In this plot we only use sources in our sample with observed $S_{850}\geq 4$\,mJy, where our sample is uniformly-selected. 

As shown in Figure \ref{fig:ncounts}, over a five arcmin diameter field, there is an excess of \submm\ sources in the clusters of \overdens\ times the expected field count down to an observed flux of $S_{850} \geq 4$\,mJy. This is a lower overdensity than in other studies of \submm\ sources in high-redshift clusters \citep[e.g.,][]{Chapman2009,Ma2015}, however we note that there was no pre-selection on star-formation activity in our cluster sample. 

Since field SMGs are known to be clustered \citep[e.g.,][]{Wilkinson2017}, to test the significance of this excess, we repeatedly select eight random regions of five arcmin diameter within the S2CLS UDS \citep{Geach2017} field survey (simulating our sample of eight clusters). We find this average overdensity ($0.50\pm0.05$)\% of the time, indicating our clusters are statistically overdense in terms of \submm\ sources, even compared to the variance in the field. 

\begin{figure}
\includegraphics[trim=0.2cm 0.2cm 0cm 1.1cm,clip,width=\columnwidth]{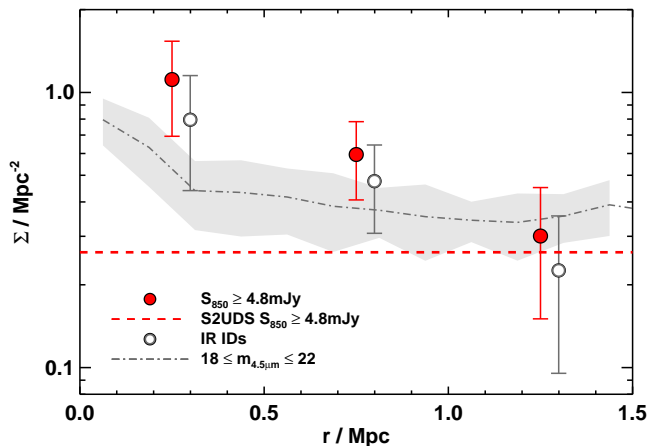}
\caption{The surface density of $850$\,\micron\ sources as a function of radius from the cluster X-ray centres. Filled points are all SCUBA-2-selected sources with $S_{850}\ge 4.8$\,mJy, which should be compared to the dashed line which shows the expected field density. We also plot the surface density of the infrared-identified SMG candidate counterparts, the majority of which are expected to lie in the clusters. The points are offset slightly for clarity. Error bars show Poisson errors on the total number of cluster members in each bin. The dot-dashed line shows the median density of IRAC-detected sources with $18\le m_{4.5\micron}\le 22$, arbitrarily scaled to match the SMG datapoints. The density of SMGs increases towards the X-ray centre of the clusters, with an overdensity above the field value of $4\pm2$ within the central $1$\,Mpc diameter. 
}
\label{fig:overdensity}
\end{figure}

To investigate the location of \submm\ sources within the clusters, in Figure \ref{fig:overdensity} we plot the excess of \submm\ sources as a function of cluster-centric radius with respect to the X-ray detected cluster core. Although our flux cut corresponds to a false detection rate of $13\%$, the majority of false detections are expected to lie towards the edges of the SCUBA-2 maps, which may bias any radial trends. We thus take a higher and more conservative flux cut of $S_{850} \geq 4.8$\,mJy, where we expect zero false detections (Figure \ref{fig:fpr}), in order to examine the radial trends. Figure \ref{fig:overdensity} shows that on average the density of \submm\ sources increases towards the X-ray centre of the clusters, with an overdensity above the field value of $4\pm2$ within 0.5\,Mpc radius. The overdensity rapidly drops at larger radii to the field density at $\gtrsim$$1$\,Mpc. This shows that the overdensity of \submm\ sources in our cluster sample is primarily within the central $\sim$$1$\,Mpc of the clusters. Integrating over all eight clusters we estimate an excess population of $\simeq 28 \pm 6$ sources above the field, most of which are within $\sim$$1$\,Mpc.

\section{Properties of submillimetre galaxies} \label{sec:properties}
We find an excess of \submm\ sources in $z\sim1$ clusters compared to the field. This excess is on average concentrated within the central $1$\,Mpc of the cluster cores. In this section we discuss the identification of the galaxies responsible for these \submm\ sources using mid-infrared data. 

\subsection{Mid-infrared identifications}
The SCUBA-2 $850$\,\micron\ effective beam FWHM is $\theta = 14.6$\,arcsec, making associations to higher-resolution data at shorter wavelengths difficult. We thus use higher-resolution infrared images from \emph{Spitzer}/MIPS at $24$\,\micron\ ($6$\,arcsec FWHM) and \emph{Herschel}/PACS at $70$ and $100$\,\micron\ ($5.2$\,arcsec and $7.7$\,arcsec FWHM, respectively), as well as our SCUBA-2 $450$\,\micron\ data ($7.5$\,arcsec FWHM), to identify probable counterparts to the \submm\ sources and obtain their infrared properties. All eight clusters have \emph{Spitzer}/IRAC $3.6$ and $4.5$\,\micron\ coverage, seven also have coverage from IRAC $5.8$ and $8.0$\,\micron, four are covered at $24$\,\micron, and six have either $70$\,\micron\ or $100$\,\micron\ data. Only one cluster does not have either $24$\,\micron\ or $70/100$\,\micron\ coverage: RCDS\,J1252. 

\begin{figure} 
\includegraphics[trim=0cm 0cm 0cm 0cm,clip,width=\columnwidth]{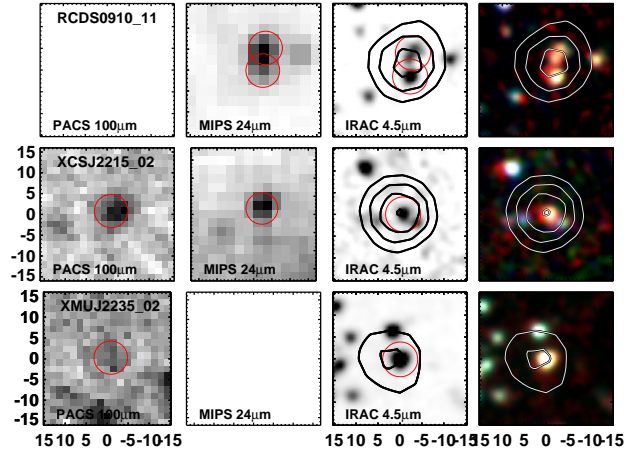}
\caption{Thumbnails of three example SCUBA-2 detections showing their infrared counterparts at $100$\,\micron, $24$\,\micron\ and $4.5$\,\micron, as well as their three-colour $3.6$\,\micron, $4.5$\,\micron, $8.0$\,\micron\ image. Contours mark the $850$\,\micron\ detections, starting at $3\sigma$ and increasing in steps of $2\sigma$. Red circles outline the positions of infrared counterparts with $p<0.05$. \emph{Top:} A bright SCUBA-2 source which resolves into two infrared counterparts, both with $p\leq0.05$, selected at $24$\,\micron\ and visible at $4.5$\,\micron. \emph{Middle:} A SCUBA-2 source with a single infrared counterpart with $p\leq0.05$, detected at all wavelengths. \emph{Bottom:} An example of a source undetected at $100$\,\micron, but selected by its IRAC colour with $p\leq0.05$. }
\label{fig:thumbs}
\end{figure}

To identify counterparts, we create catalogues of infrared sources using SExtractor \citep{SExtractor} on the \emph{Spitzer}/IRAC $4.5$\,\micron\ images and measuring fluxes at the resulting positions at $24$, $70$, $100$, and $450$\,\micron, where available. Some example mid-infrared thumbnails for three \submm\ sources are shown in Figure \ref{fig:thumbs}. We then calculate a corrected-Poissonian probability \emph{p}-value \citep{Downes1986, Dunlop1989} for all $24$\,\micron, $70$\,\micron\ and $100$\,\micron-detected sources within the SCUBA-2 cluster maps. 

The probability that a given infrared source is associated with an $850$\,\micron\ source is a function of magnitude and separation. For each infrared/\submm\ source pairing, 
\begin{equation}
\qquad P^* = \pi r^2 N_{m<M}
\end{equation}
where $r$ is the offset between the infrared and \submm\ sources and $N_{m<M}$ is the number density of infrared sources in the field which have a magnitude $m$ brighter than the magnitude of the infrared source, $M$. Given a value of $P^*$, we can derive the probability that the infrared source is a chance alignment with the $850$\,\micron\ source, $p=(1-\exp[-E])$, where $E$ is given by:

\begin{align}
\qquad E = P^*  \qquad P^* > P_c \nonumber \\ 
\qquad E = P^*(1+\log[P_c/P^*])  \qquad P* \le P_c
\end{align}

\begin{figure} 
\begin{centering}
\includegraphics[trim=0cm 0cm 0cm 0.5cm,clip,width=0.9\columnwidth]{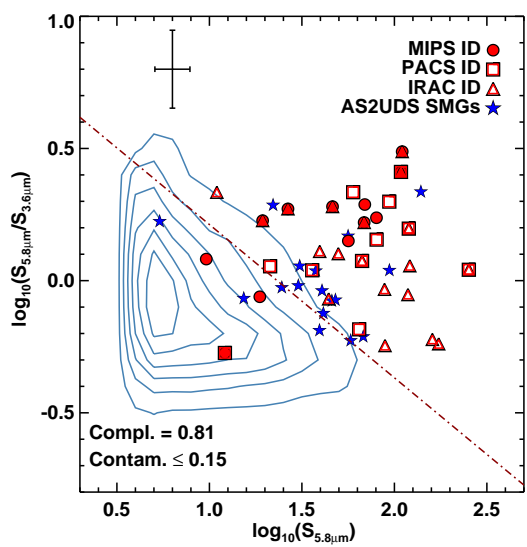}
\caption{The \emph{Spitzer}/IRAC properties of infrared counterparts to SCUBA-2 $850$\,\micron\ sources in our cluster survey. The contours show the distribution of field galaxies in the UKIDSS Ultra Deep Survey (UDS) field (Almaini et al., in preparation). The dot-dashed line shows the \emph{scikit learn}-derived separation between $z\sim0.8$--$1.6$ AS2UDS SMGs \citep{Stach2019} and field galaxies, used to select IRAC counterparts to $850$\,\micron\ sources with no MIPS or PACS coverage. This shows that using this selection and a matching probability$p\leq0.05$ results in a sample with similar properties to known SMGs. A typical error bar for the candidate SMG counterparts is shown in the top left. }
\label{fig:IRselection}
\end{centering}
\end{figure}

\noindent $P_c$ is the critical Poission probability level, $P_c = \pi r^2_s N_T$, where $N_T$ is the total surface density of all detected infrared sources and $r_s$ is the search radius (here we use $7.3$\,arcsec, the half-width at half maximum of the SCUBA-2 $850$\,\micron\ beam). 
We identify infrared counterparts to \submm\ sources (Figure \ref{fig:thumbs}) if their \emph{p}-value is $p \le 0.05$, based on the results from \citet{Hodge2013} \citep[see also,][]{An2018}. We select $23$ infrared counterparts to $15/83$ \submm\ sources. The colour-magnitude diagram for these candidate counterparts is shown in Figure \ref{fig:IRselection}. 

The cluster RCDS\,J1252 has no coverage by MIPS or PACS. We thus use the IRAC properties of the infrared counterpart SMGs in the other clusters to determine probable IRAC counterparts in RCDS\,J1252. We use the $3.6$\,\micron\ and $5.8$\,\micron\ fluxes from a  $0.8<z<1.6$ field sample from the UKIDSS/UDS (Almaini et al., in preparation), and $0.8<z<1.6$ SMGs from the ALMA/SCUBA-2 UDS survey \citep[AS2UDS;][]{Stach2018,Stach2019} plotted in Figure \ref{fig:IRselection} to determine an infrared colour/magnitude selection for likely $z\sim0.8$--$1.6$ SMGs. Following \citet{An2018}, we apply a linear support vector classification using the {\sc python} package \emph{scikit-learn}\footnote{\href{http://scikit-learn.org}{http://scikit-learn.org}} \citep{scikit-learn} to derive the optimal SMG selection:
\begin{equation} \label{eqn:iracsel}
\qquad  \log_{10}(S_{5.8}/S_{3.6}) > 1.099 - 0.835\times \log_{10}(S_{5.8}) 
\end{equation}

\noindent We then select as SMG counterparts any sources which satisfy Equation \ref{eqn:iracsel} (``IRAC colour-selected sources") and have \emph{p}-values (using the $3.6$\,\micron\ magnitudes) $p \le 0.05$. We find $19$ IRAC colour-selected counterparts, nine of which also have a $24$\,\micron\ and/or a $70/100$\,\micron\ detection. We test this method by randomising the positions of all IRAC colour-selected sources in each cluster field and re-measuring their \emph{p}-values. We select a source with $p \le 0.05$ in five percent of the randomisations. We therefore expect to find one IRAC colour-selected counterpart to four SCUBA-2 sources due to random alignments, compared to the $19$ candidate counterparts that we identify. This is an upper limit at it does not take into account any additional information from $24$\,\micron, $70$\,\micron\ or $100$\,\micron\ detections. 

\citet{Hodge2013} showed that by using mid-infrared detections, counterparts to single-dish \submm\ sources are correct in $80\%$ of cases, but are only recovered in $45\%$ of \submm\ sources. We identify counterparts to $33\%$ ($26/83$) of the \submm\ sources in our sample which is consistent with the findings of \citet{Hodge2013} and our estimate of the likely number of $850$\,\micron\ sources which are associated with the cluster overdensities in Section \ref{sec:ncounts}. 
Our cluster sample is at a lower redshift than the average of the sample from \citet{Hodge2013} \citep[$z\sim2.5$;][]{DaCunha2015}, meaning the $K$-correction at $24$\,\micron\ and $70/100$\,\micron\ is smaller and thus making it easier to detect mid-infrared counterparts that are candidate cluster members. We therefore expect our counterparts to be $\gtrsim$$80\%$ accurate. 

To summarise, we select SMGs as any sources with $p\leq 0.05$ which have a $24$\,\micron\ counterpart and/or a $70/100$\,\micron\ counterpart and/or an IRAC colour-selected source. Table \ref{table:IR} lists the properties of all the infrared counterpart SMGs as well as the method used to identify them. 
In total we find $26/83$ SCUBA-2 sources have at least one infrared-selected counterpart, with $34$ infrared-selected SMGs in total from the $26$ SCUBA-2 sources. $19/83$ SCUBA-2 sources have IRAC colour-selected counterparts, nine of which also have a $24$\,\micron\ and/or a $70/100$\,\micron\  detection. $9/83$ and $9/83$ sources have $24$\,\micron\ and $70/100$\,\micron-selected counterparts, respectively.  

$57/83$ SCUBA-2 sources do not have any infrared counterpart assigned to them. This gives a (completeness-corrected) number density of $\sim$$1300$\,deg$^{-2}$ for the SCUBA-2 sources brighter than $4$\,mJy which lack counterparts, consistent with the expected surface density of the field population (Figure \ref{fig:ncounts}), which are typically at higher redshifts \citep[$z\sim2.5$;][]{Danielson2017,Stach2019}. These sources are probably background field SMGs and not cluster members, although spectroscopic redshift information is required to confirm this. 

Of our $34$ infrared-selected SMGs we expect the majority to be cluster members due to the smaller $K$-correction in the infrared at $z\sim0.8$--$1.6$, compared to the average redshift of SMGs ($z\sim2.5$). Future spectroscopic observations of these targets will be able to confirm their cluster membership, and constrain their relative velocities within the cluster.

We find that $7/26$ \submm\ sources which have robust infrared identifications have more than one counterpart with $p\leq0.05$. If all of these counterparts are SMGs this suggests a multiplicity rate for the single-dish sources of $30\pm10\%$. This is slightly higher than the rate in field surveys, which have a multiple fraction of $\gtrsim15\%$ for $S_{850}\geq3.5$\,mJy \citep{Stach2018}. However, we stress that to reliably identify SMG counterparts to \submm\ sources higher-resolution \submm\ observations, such as from ALMA or the Sub-Millimeter Array (SMA), are crucial. This is particularly true for crowded fields such as the cluster cores in this sample, as the density of potential counterparts is much higher. Indeed, \citet{Stach2017} used ALMA observations to resolve four single-dish \submm\ sources into $14$ separate SMGs in the core of XCS\,J2215. 

\subsection{Testing cluster membership} \label{sec:evol}
Most of the clusters in this study have spectroscopic coverage in the optical or near-infrared. We have searched for any archival spectroscopic redshifts for our candidate cluster members and find two matches: RX\,J0849\_02a and XCS\,J2215\_06a. The archival redshift for RX\,J0849\_02a places it at $z=1.589$, which indicates that this is a background source and not a cluster member. The source in XCS\,J2215 is a spectroscopically confirmed cluster member \citep{Stach2017}. We also note that there are a further $11$ spectroscopically-confirmed, \submm-detected cluster members from \citet{Stach2017} and \citet{Hayashi2017} which are not selected in our sample because they have $S_{850} < 4$\,mJy. To confirm the membership of our sample, future deep near-infrared or \submm\ spectroscopy is required.

\begin{figure}
\includegraphics[trim=0.3cm 0cm 0cm 0.7cm,clip,width=\columnwidth]{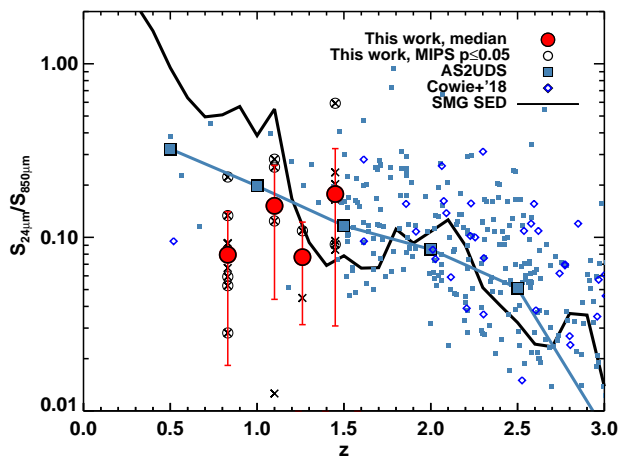}
\caption{The evolution of the $S_{850}$/$S_{24}$ flux ratio with redshift for our cluster SMGs and samples of field SMGs from AS2UDS \citep[][Dudzevi{\v c}i{\=u}t{\.e} et al., in preparation]{Stach2018} and GOODS-S \citep{Cowie2018}. Where MIPS coverage is available, all SMGs are detected at $24$\,\micron. Small black crosses without circles have been selected via their $70/100$\,\micron\ or IRAC colour. We plot the median and scatter for individual clusters and the median field flux ratio in redshift bins. The black solid line is a composite SMG SED \citep{Swinbank2014}. Albeit with large scatter between potential cluster members, overall cluster SMGs appear to have similar colours to field SMGs. The SMGs in RX\,J0152 at $z=0.83$ have a median ratio a factor of two lower than the general SMG population at $z\simeq0.8$, which may indicate contamination from higher-redshift SMGs.
}
\label{fig:850v24vz}
\end{figure}

Due to the negative \emph{K}-correction at \submm\ wavelengths, the ratio of $24$\,\micron\ flux density to $850$\,\micron\ flux density is expected to decrease towards higher redshifts \citep[e.g.,][]{Cowie2018}. Four of the eight clusters have MIPS $24$\,\micron\ coverage sensitive enough to detect flux ratios down to $S_{24}/S_{850}\sim0.02$. Within these four clusters, $12$ sources have a MIPS identification with $p\leq 0.05$ and a further $13$ have measurable $24$\,\micron\ fluxes (Table \ref{table:IR}). In Figure \ref{fig:850v24vz} we plot the evolution of the $24$\,\micron/$850$\,\micron\ flux ratio with redshift for SMGs from the AS2UDS survey \citep{Stach2019} and from the GOODS-S field \citep{Cowie2018} and compare to those for our sample of four clusters. The cluster sample on average has $24$\,\micron/$850$\,\micron\ flux consistent with the field at $1\lesssim z\lesssim1.5$, albeit with a large scatter between potential cluster members. This is further evidence that by selecting \submm\ sources with infrared counterparts, we are selecting probable cluster member SMGs rather than background sources. 

The SMGs in RX\,J0152 at $z=0.83$ have a median $S_{24}/S_{850}$ ratio a factor of two lower than the field SMG population at $z\sim0.8$. This cluster has been extensively studied and has been shown to have an irregular structure and strongly-lensing core. We discuss RX\,J0152 further in section \ref{sec:sfrs}. The lower flux ratios, however, may indicate that some of the SMGs we observe in this cluster field are lensed background galaxies, with colours more consistent with $2<z<3$ SMGs.

\section{Results and Discussion} \label{sec:discussion}
\subsection{Radial overdensity} \label{sec:overdensity}
In local clusters, star-forming galaxies are preferentially located on the outskirts of these massive structures, whereas the core is populated by passive galaxies \citep[e.g.,][]{vonderLinden2010,YPeng2010}. The location of star-forming members provides indicators for the formation and quenching mechanisms of cluster galaxies. For example, galaxies falling into the dense intra-cluster medium may have their cold gas stripped and thus cease forming stars \citep[e.g.,][]{Gunn1972,Jaffe2018}. Conversely, interactions and mergers between gas-rich galaxies may cause starburst events \citep[e.g.,][]{Mihos1994,Kocevski2011}. 

Using $24$\,\micron\ and $70/100$\,\micron\ counterparts we expect to predominantly select SMG members of the $z\sim1$ clusters, rather than background interlopers. In Figure \ref{fig:overdensity} we showed that the density of \submm\ sources increases towards the X-ray-defined centre of the clusters, with an overdensity above the field value of $4\pm2$ within $1$\,Mpc. We also show SMGs for which we identify infrared counterparts. The infrared counterparts follow the same overall trend as the \submm\ detections, although with lower significance. 

The increase in density of our candidate SMGs near the X-ray centre of the clusters suggests that the candidate cluster SMGs lie within the central $\sim$$500$\,kpc of the structures. The short gas-consumption timescale of SMGs \citep[typically $10^7$--$10^8$\,yr; ][]{Bothwell2013} means that they are unlikely to have moved far from the environment where the intense star-formation event began ($\sim$$0.01$--$0.1$\,Mpc, assuming a velocity of $1000$\,km\,s$^{-1}$). This means that the star-formation event was likely triggered within the central $\sim1$\,Mpc core of these massive clusters. 

A number of studies have suggested that the triggering mechanism for SMGs may be interactions or mergers \citep[e.g.,][]{Swinbank2006,Ivison2007,Engel2010,Chen2015}. The overdensity of dust-obscured star formation within the cores of these $z\sim1$ clusters therefore may suggest an overdensity of gas-rich mergers between cluster members at radii $<500$\,kpc.

\subsection{Total cluster star-formation rates} \label{sec:sfrs}

\begin{figure} 
\includegraphics[width=\columnwidth]{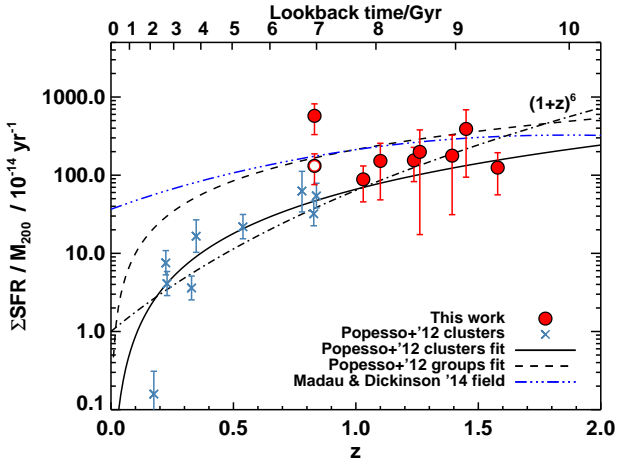}
\caption{The total star-formation rate per cluster mass ($M_{200}$) as a function of redshift. The small blue points and lines show previous work at $z<1$ \citep{Popesso2012}. The open symbol shows the value of RX\,J0152 if we use the cluster mass from \citet{Popesso2012}. The ratio is still higher than their value due to our measured higher star-formation rate, which may be contaminated by background sources (see Section \ref{sec:evol}). The \submm\ star-formation rates of the cluster sample are consistent with a continuation of the lower-redshift trend within the scatter, but can also be fit by a steeper relation $\sim$$(1+z)^6$.  }
\label{fig:Popesso}
\end{figure}

Although in the local Universe there is a clear trend of lower star-formation rates in denser environments \citep[e.g,][]{Kauffmann2004}, at higher redshifts there are indications that this trend reverses at the epoch of cluster galaxy formation \citep[e.g.,][but see also \citealt{Quadri2012,Ziparo2014,Muldrew2018}]{Elbaz2007,Tran2010,Elbaz2011,Brodwin2013}. In addition, previous studies of individual clusters suggested a rapid evolution in the mass-normalised star-formation rate of $\propto(1+z)^\gamma$, with $\gamma=6$--$7$ \citep[e.g.,][]{Kodama2004,Geach2006,Koyama2010,Koyama2011,Shimakawa2014,Smail2014}, compared to $\gamma=3$--$4$ for field galaxies \citep[e.g.,][]{Ilbert2015}. Here we examine the mass-normalised total star-formation rate of our cluster candidate sample compared to lower-redshift results. 

To calculate cluster masses we use their X-ray temperatures ($T_{\rm X}$) from the literature \citep[Table \ref{table:clusters};][]{Stanford2001,Stanford2002,Maughan2003,Rosati2004,Mullis2005,Stanford2006,Branchesi2007,Tozzi2015} and the relation between $T_{\rm X}$ and  $M_{500}$ from \citet{Kettula2013}. We then convert to $M_{200}$ assuming the density profile from \citet{NFW} with a concentration value of $5$ \citep{Bullock2001}. Star-formation rates were calculated using a conversion between the measured $850$\,\micron\ flux and star-formation rate:
\begin{equation}
\qquad \log_{10}(SFR) = (0.87\pm0.06) \times \log_{10}(S_{850}) + (1.85\pm0.04)
\end{equation}

\noindent calculated from fitting the star-formation rate derived using {\sc magphys} \citep{DaCunha2008} on the full spectral energy distribution for the $45$ SMGs at $z_{\rm phot}=0.8$--$1.6$ drawn from a full sample of over $700$ SMGs in the AS2UDS survey \citep[][Dudzevi{\v c}i{\=u}t{\.e} et al., in preparation]{Stach2018,Stach2019}. This scaling relation is derived between observed $850$\,\micron\ flux and far-infrared-derived star-formation rate and has a dispersion of $0.27$\,dex. 

Figure \ref{fig:Popesso} shows the integrated star-formation rate normalised by cluster mass. We see that our clusters at $z=0.8$--$1.6$ are consistent with the overall trend of higher mass-normalised star-formation rate at higher redshifts. Our sample is consistent with a continuation of the trend found in \citet{Popesso2012} or the $\sim$$(1+z)^6$ evolution suggested by intermediate-redshift studies of H$\alpha$ emitters \citep[e.g.][]{Kodama2004,Koyama2010,Koyama2011}. Our current data are unable to distinguish between these trends.

One of the clusters in this work, RX\,J0152, was also studied by \citet{Popesso2012}. We find an integrated star-formation rate to cluster mass ratio a factor of $16$ times larger than that study. This is due to both a factor of four times lower cluster mass estimate and the factor of four times higher measured star-formation rates in our study. \citet{Popesso2012} measured cluster masses using cluster members' velocity dispersions, whereas we convert the X-ray temperature as above. In Figure \ref{fig:Popesso} we show as an open symbol the value if we instead adopt the cluster mass listed in \citet{Popesso2012}. 
In addition, our measured star-formation rates are higher than those listed in \citet{Popesso2012}. This may indicate that the large effective beam of SCUBA-2 means that the $850$\,\micron\ flux measurements are potentially contaminated by background sources, boosting the measured star-formation rates. The data points in Figure \ref{fig:Popesso} may therefore be considered upper limits, however we note that we have taken a conservative flux cut and thus may also be missing fainter cluster members, which would increase the total star-formation rates. 

Previous studies of RX\,J0152 have revealed a double-core system, indicative of an early-stage cluster-to-cluster merger \citep[e.g.,][]{Rosati1998,Maughan2003,Tanaka2006}. The central regions of RX\,J0152 are also known to be forming strong lensing multiple images of background systems \citep{Umetsu2005,Acebron2018}. None of our \submm\ sources are identified as lensed by recent strong lensing studies \citep{Acebron2018}, however if some of the detected \submm\ sources in the cluster are actually lensed background sources then this may increase our measured star-formation rate, and may explain the lower $S_{24}/S_{850}$ ratios in Section \ref{sec:evol}. Further spectroscopic observations are required to determine the cluster membership of the observed SMGs. However, we note that if we remove RX\,J0152 from our sample none of our results qualitatively change. 

In Figure \ref{fig:overdensity} we showed that the overdensity of SMGs in our cluster sample is strongest within the central $0.5$\,Mpc. Two clusters in our sample, XCS\,J2215, and RX\,J0152 (at $z=1.45$ and $z=0.83$, respectively) have bi-modal cores, indicating they are likely undergoing a cluster-to-cluster merger \citep{Maughan2003,Stanford2006,Hilton2010}. In Figure \ref{fig:Popesso}, XCS\,J2215 and RX\,J0152 are the clusters with the highest star-formation rate densities. This may be hinting that cluster-to-cluster mergers may be responsible for triggering extreme star-formation activity within the resulting system's core. 

We now examine whether the overdensity of star-forming galaxies in the cores of these clusters indicates a reversal in the local star-formation rate-density relation. The median normalised star-formation rate for the central $\sim2.5$\,Mpc of the $z=0.8$--$1.6$ clusters\footnote{Including RX\,J0152 the median is $\Sigma SFR/M_{\rm cl} = 200_{-60}^{+200} \times 10^{-14}$\,yr$^{-1}$} is $\Sigma SFR/M_{\rm cl} = 180_{-50}^{+20} \times 10^{-14}$\,yr$^{-1}$. To calculate the equivalent value for field galaxies we follow \citet{Popesso2012} and adopt the cosmic star-formation rate density from \citet{Madau2014} and divide it by the mean comoving density of the Universe ($\Omega_M \times \rho_{\rm crit}$, where $\rho_{\rm crit}$ is the critical density of the Universe) to get $SFR/M = 270_{-60}^{+30} \times 10^{-14}$\,yr$^{-1}$. The normalised star-formation rate for the candidate cluster sample is lower than the field by a factor of $1.5\pm0.3$.  This suggests that there is no evidence for a systematic reversal in the star-formation rate-density relation on $2.5$\,Mpc scales up to $z=1.6$, however, we note that we have taken conservative flux cuts for our SMG sample. ALMA observations of XCS\,J2215 revealed $14$ SMGs within the central $500$\,kpc of the cluster core \citep{Hayashi2017,Stach2017}. Future deep observations to select fainter \submm\ sources, and spectral analysis to confirm their cluster membership may uncover more SMGs and thus higher total star-formation rates within these cluster cores.

\begin{figure} 
\includegraphics[trim=0.7cm 0cm 0cm 0.5cm,clip,width=\columnwidth]{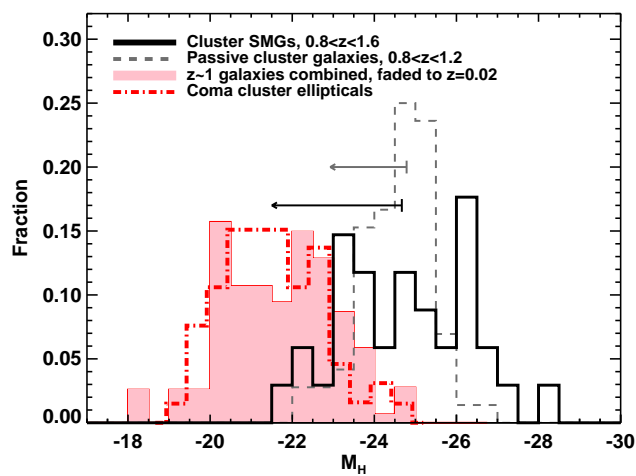}
\caption{Comparison of the distributions of absolute $H$-band magnitude for our cluster SMG sample, a spectroscopic $z\sim1$ passive cluster sample \citep{Muzzin2009,Demarco2010} and elliptical galaxies from the Coma cluster. The shaded histogram shows the distribution of the SMGs and passive samples evolved to $z=0.023$ (shown by the arrows) and combined. For this we assume that the SMG starbursts we observe are $100$\,Myr in duration and on average we observe the SMGs halfway through the starburst. If we instead consider a $50$\,Myr or $200$\,Myr burst the SMG distribution shifts by $+0.06$\,mag and $-0.5$\,mag, respectively. A two-sided Kolmogorov-Smirnov test comparing the combined sample to the Coma distribution gives $p=0.7$. The passive population in $z\sim0$ clusters is consistent with forming at $z\ge1$ through a series of dusty starbursts. 
}
\label{fig:MH}
\end{figure}

\subsection{Future evolution}
SMGs have been suggested as the progenitors of local spheroidal galaxies. We therefore examine the future evolution of our cluster candidate sample to see whether their luminosity distribution is consistent with local passive cluster galaxies. To ``evolve" our population to $z=0$ we follow the method of \citet{Simpson2014} and calculate the expected change in rest-frame $H$-band magnitude (observed-frame $3.6$--$4.5$\,\micron) from $z\sim1.2$\footnote{We calculate the expected change in $H$-band magnitude for each cluster individually to account for the different redshifts.} to $z=0$. This method uses \citet{Bruzual2003} simple stellar population models to determine the $H$-band luminosity halfway through a $100$\,Myr burst (taken as the typical lifetime of an SMG) at $z\sim1$ and predict its evolution to the present day. We also run these models with $50$\,Myr and $200$\,Myr bursts and show the offset in $H$-band magnitude for these as vectors in Figure \ref{fig:MH}. This is a simple model and assumes that the pre-burst stellar population's luminosity contribution is negligible and that each SMG only goes through a single burst phase and does not subsequently accrete significant stellar mass through dry mergers. 
Figure \ref{fig:MH} shows the absolute $H$-band magnitudes of the candidate cluster SMGs, faded to $z=0.023$, compared to present-day elliptical galaxies in the Coma cluster \citep[][R. Smith, private communication]{Smith2009,Hainline2011}. The SMG distribution has a median $H$-band magnitude of $M_H=-21.5\pm0.2$, consistent with that of present-day cluster ellipticals ($M_H=-21.2\pm0.3$), suggesting these candidate cluster SMGs could evolve into passive ellipticals by $z=0$.

To estimate whether the number of SMGs we observe in our cluster sample at $z\sim1$ are sufficient to explain the observed passive population in clusters at $z=0$ we estimate the infall rate and duty cycle of the SMG phase to correct our apparent numbers of SMGs in $z\sim1.2$ clusters to the expected present-day descendants and also add in those galaxies which are already passive at $z\sim1.2$. To do this we use the implementation of the \citet{Bower2006} galaxy-formation recipe in the Millennium Simulation \citep{Springel2005}. We use the most massive halo in the simulation (which has a mass of $\sim$$1\times10^{14}$\,\Msun\ at $1\lesssim z\lesssim 2$, roughly matching the expected masses of our cluster sample) and a spherical search radius of $1.5$\,Mpc, matching our SCUBA-2 survey field. We searched the simulation snapshots between $z=0.91$ and $z=1.84$ for galaxies with cold gas masses of $>$$1\times10^{10}$\,\Msun\ lying within this sphere. We then identify unique entries in the output for each of these gas-rich galaxies, on the expectation that the gas reservoir will be quickly exhausted by the intense star-formation events we are searching for. There are a total of $40$ such galaxies found over a duration of $2.7$\,Gyr, with an average of $1.5\pm0.7$ SMGs in any given $100$\,Myr snapshot. Given that the average SMG phase lasts approximately $100$\,Myr, this is the expected number of observable SMGs per cluster at these redshifts. The median of our sample is two SMGs per cluster. Thus our approximation, although crude, is in agreement with our observations. This suggests a correction factor to account for all such \submm-bright galaxies accreted onto a typical cluster since $z\sim2$ of $\sim$$25\pm7$.  
	
A similar calculation using the average cluster mass growth and stellar mass function taken from \citet{Muldrew2015} suggests that we would expect to observe $1$--$3$ SMGs per cluster and a correction factor of $\sim$$10$. Our estimated correction factor may therefore change by a factor of two. We have tested our analysis with both correction factors and find our conclusions do not qualitatively differ.  

Although the median of our SMG sample is consistent with that of present-day cluster ellipticals, by $z\sim1$ most clusters already have a population of passive galaxies \citep[e.g.,][]{Eisenhardt2008}. In Figure \ref{fig:MH} we therefore combine our SMG sample with a $z\sim1$ spectroscopic sample from \citet{Muzzin2009} and \citet{Demarco2010}. This sample is incomplete at faint magnitudes ($M_{\rm H} \geq -24$). We have tested the effect this may have on our conclusions by  multiplying the faint end of the spectroscopic distribution by a factor of $100$ before combining them with the SMG sample. We find that our conclusions do not qualitatively change. We scale the number of SMGs observed by a factor of $25$, as calculated above, to account for the duty cycle of the SMG phase, fade them to $z\sim0$ and then combine the distribution with that of the faded $z\sim1$ passive galaxies. As shown in Figure \ref{fig:MH}, the combined SMG$+$passive distribution is similar to that of the Coma ellipticals in terms of median luminosity and width; a two-sided Kolmogorov-Smirnov test gives $p=0.7$. This therefore suggests that the passive population in $z\sim0$ clusters is consistent with having formed all of its stars at $z\ge1$ through a series of dusty starbursts.

\section{Conclusions} \label{sec:concl}
We have observed a sample of eight massive galaxy clusters at $0.8<z<1.6$ with SCUBA-2 at $850$\,\micron. We select $83$ \submm\ sources with either $S/N \ge 4$ or $S_{850}\ge 4$\,mJy within $2.5$\,arcmin radii of the cluster cores ($1.2$\,Mpc at $z=1.2$). We find an overdensity of \submm\ sources of a factor of \overdens\ over the expected field density. This overdensity is mostly concentrated within $\sim$$500$\,kpc around the X-ray detected cluster cores: a factor of $4\pm2$ overdense within $1$\,Mpc diameter, suggesting there is ongoing dusty star formation in the centres of massive clusters at $z\sim1$.  

We use higher-resolution infrared images to select likely cluster member SMGs. We match $26$ of the low-resolution \submm\ sources to $34$ likely infrared counterparts and examine their multiwavelength properties. The remaining $56$ \submm\ sources have a number density consistent with the field population and are therefore expected to be background field SMGs at higher redshifts. We find that the total amount of star formation, normalised by cluster mass, increases out to $z\sim1.5$ and is consistent with a more rapid evolution ($\sim$$(1+z)^6$) than the $\sim$$(1+z)^4$ trend from the field. Even with this rapid evolution, the mass-normalised star-formation rate for clusters at $1<z<1.6$ is lower than the field by a factor of $1.5\pm0.3$. We therefore find no evidence with our current data of a reversal of the local star-formation rate-density relation in the most massive X-ray-detected clusters at $z\sim0.8$--$1.6$. 

Finally, we use a simple model to predict the $H$-band luminosities of our candidate cluster SMGs evolved to the present day and compare this to both local cluster ellipticals, and the population of cluster galaxies which are already passive by $z\sim1$. We find that the evolved distribution of $M_H$ from our star-forming cluster sample is consistent with that of faint ($M_H\gtrsim-22$) passive elliptical galaxies in the Coma cluster. Combining the passive cluster population at $z\sim1$ with the SMG sample we can reproduce the expected cluster population at $z=0$. This suggests that the majority of the passive population in $z\sim0$ clusters are consistent with having formed at $z\gtrsim1$--$1.5$ through an extreme, dust-obscured starburst event.

\section*{Acknowledgments}
The authors would like to thank the anonymous referee for their comments which improved the flow and content of this paper. 
The authors would also like to thank the following people for useful discussions and help with the survey: James Simpson, Renske Smit, C.\ J.\ Ma, Alex Karim, John Stott, Ken Tadaki, Masayuki Tanaka, Simona Mei and John Blakeslee. 
EAC and IRS acknowledge support from the ERC Advanced Investigator Grant DUSTYGAL (321334) and STFC (ST/P000541/1). 
We also acknowledge STFC support to the UK consortium members of EAO (ST/M007634/1, ST/M003019/1 and ST/N005856/1). 
APT acknowledges support from STFC (ST/P000649/1). 

This work makes use of data from JCMT project IDs M13AU029, M13BU010, M15BI006, M16AP053, and M16BP080. 
The James Clerk Maxwell Telescope is operated by the East Asian Observatory on behalf of The National Astronomical Observatory of Japan; Academia Sinica Institute of Astronomy and Astrophysics; the Korea Astronomy and Space Science Institute; the Operation, Maintenance and Upgrading Fund for Astronomical Telescopes and Facility Instruments, budgeted from the Ministry of Finance (MOF) of China and administrated by the Chinese Academy of Sciences (CAS), as well as the National Key R\&D Program of China (No.\ 2017YFA0402700). Additional funding support is provided by the Science and Technology Facilities Council of the United Kingdom and participating universities in the United Kingdom and Canada. 

This work is based in part on archival data obtained with the \emph{Spitzer Space Telescope} and the NASA/IPAC Extragalactic Database (NED), which are operated by the Jet Propulsion Laboratory, California Institute of Technology under a contract with the National Aeronautics and Space Administration. 
This research also used the facilities of the Canadian Astronomy Data Centre operated by the National Research Council of Canada with the support of the Canadian Space Agency. 

\bibliographystyle{mnras}
\bibliography{SMGs.bib} 

\appendix \label{section:appendix}
\begin{table}
\setcounter{table}{1}
\caption{Table of identifications and infrared properties of SCUBA-2-identified \submm\ sources. All sources with S/N\,$\geq4$ were selected, as well as those with  S/N\,$<4$ but $S_{850\micron}>4$\,mJy. Fluxes given are observed peak fluxes for each source in the corresponding map. }
\begin{tabular}{ lrrrr}
\hline
\multicolumn{1}{c}{ID} &
\multicolumn{1}{c}{R.A.} &
\multicolumn{1}{c}{Decl.} &
\multicolumn{1}{c}{$S_{850}$} &
\multicolumn{1}{c}{$S_{450}$} \\
\multicolumn{1}{c}{} &
\multicolumn{2}{c}{[J2000]} &
\multicolumn{1}{c}{[mJy]} &
\multicolumn{1}{c}{[mJy]} \\
\hline
\hline
RXJ0152\_01 & 01:52:44.02 & $-$13:58:53.0 &  4.0$\pm$0.6  &   20$\pm$5    \\
RXJ0152\_02 & 01:52:33.58 & $-$13:58:13.0 &  3.8$\pm$0.6  &    7$\pm$5    \\
RXJ0152\_03 & 01:52:42.10 & $-$13:58:05.0 &  7.6$\pm$0.5  &   22$\pm$4    \\
RXJ0152\_04 & 01:52:44.02 & $-$13:57:37.0 &  3.4$\pm$0.5  &   16$\pm$4    \\
RXJ0152\_05 & 01:52:34.41 & $-$13:57:41.0 &  2.3$\pm$0.5  &  $<$4  \\
RXJ0152\_06 & 01:52:37.98 & $-$13:57:41.0 &  2.0$\pm$0.5  &    7$\pm$4    \\
RXJ0152\_07 & 01:52:40.18 & $-$13:57:09.0 &  4.1$\pm$0.5  &   10$\pm$4    \\
RXJ0152\_08 & 01:52:42.65 & $-$13:57:01.0 &  4.3$\pm$0.5  &  $<$4  \\
RXJ0152\_09 & 01:52:49.52 & $-$13:56:57.0 &  7.1$\pm$0.7  &    6$\pm$5    \\
RXJ0152\_10 & 01:52:33.03 & $-$13:57:05.0 &  4.0$\pm$0.6  &   17$\pm$5    \\
RXJ0152\_11 & 01:52:51.17 & $-$13:56:33.0 &  3.5$\pm$0.7  &   16$\pm$5    \\
RXJ0152\_12 & 01:52:44.30 & $-$13:56:17.0 &  2.6$\pm$0.6  &   12$\pm$5    \\
RXJ0152\_13 & 01:52:41.00 & $-$13:55:57.0 & 10.6$\pm$0.6  &   27$\pm$5    \\
RXJ0152\_14 & 01:52:45.67 & $-$13:55:09.0 &  5.2$\pm$0.9  &   13$\pm$6    \\
WARP1415\_01 & 14:15:10.77 & $+$36:10:59.0 &  3.5$\pm$0.5  &   10$\pm$3    \\
WARP1415\_02 & 14:15:10.77 & $+$36:10:19.0 &  3.4$\pm$0.6  &    8$\pm$4    \\
WARP1415\_03 & 14:15:10.77 & $+$36:10:39.0 &  2.2$\pm$0.6  &    7$\pm$3    \\
WARP1415\_04 & 14:15:23.99 & $+$36:11:07.0 &  4.3$\pm$0.7  &    5$\pm$4    \\
WARP1415\_05 & 14:15:12.09 & $+$36:11:23.0 & 19.2$\pm$0.5  &   41$\pm$3    \\
WARP1415\_06 & 14:15:15.40 & $+$36:11:47.0 &  2.9$\pm$0.5  &   11$\pm$3    \\
WARP1415\_07 & 14:15:13.74 & $+$36:12:11.0 &  2.3$\pm$0.5  &    4$\pm$3    \\
WARP1415\_08 & 14:15:23.66 & $+$36:12:39.0 &  4.7$\pm$0.7  &    6$\pm$4    \\
WARP1415\_09 & 14:15:05.15 & $+$36:12:39.0 &  5.1$\pm$0.5  &   11$\pm$3    \\
WARP1415\_10 & 14:15:11.43 & $+$36:13:03.0 &  2.2$\pm$0.5  &    6$\pm$3    \\
WARP1415\_11 & 14:15:17.71 & $+$36:13:43.0 &  4.8$\pm$0.6  &   15$\pm$4    \\
WARP1415\_12 & 14:15:08.79 & $+$36:14:47.0 &  7.9$\pm$0.6  &   23$\pm$4    \\
RCDS0910\_01 & 09:10:40.88 & $+$54:20:44.0 &  3.8$\pm$0.7  &   16$\pm$4    \\
RCDS0910\_02 & 09:10:52.32 & $+$54:21:16.0 &  3.3$\pm$0.6  &    9$\pm$4    \\
RCDS0910\_03 & 09:10:48.66 & $+$54:21:04.0 &  2.6$\pm$0.6  &    5$\pm$4    \\
RCDS0910\_04 & 09:10:45.46 & $+$54:21:24.0 &  6.0$\pm$0.6  &   16$\pm$4    \\
RCDS0910\_05 & 09:10:54.61 & $+$54:21:44.0 &  3.4$\pm$0.6  &   11$\pm$4    \\
RCDS0910\_06 & 09:10:39.96 & $+$54:21:44.0 &  3.1$\pm$0.6  &   12$\pm$4    \\
RCDS0910\_07 & 09:10:48.20 & $+$54:21:44.0 &  2.6$\pm$0.6  &    8$\pm$4    \\
RCDS0910\_08 & 09:11:06.97 & $+$54:22:07.9 &  5.0$\pm$0.7  &   16$\pm$5    \\
RCDS0910\_09 & 09:11:05.14 & $+$54:22:15.9 &  4.4$\pm$0.7  &   16$\pm$5    \\
RCDS0910\_10 & 09:10:58.73 & $+$54:22:07.9 &  4.3$\pm$0.7  &    7$\pm$4    \\
RCDS0910\_11 & 09:10:55.07 & $+$54:22:20.0 &  5.5$\pm$0.6  &   21$\pm$4    \\
RCDS0910\_12 & 09:10:49.58 & $+$54:22:40.0 &  2.6$\pm$0.6  &    9$\pm$4    \\
RCDS0910\_13 & 09:10:56.91 & $+$54:23:04.0 &  3.9$\pm$0.7  &   15$\pm$4    \\
RCDS0910\_14 & 09:10:50.50 & $+$54:23:20.0 &  2.8$\pm$0.6  &   15$\pm$4    \\
RCDS1252\_01 & 12:52:54.39 & $-$29:27:57.0 &  2.7$\pm$0.6  &   17$\pm$5    \\
RCDS1252\_02 & 12:52:47.96 & $-$29:27:53.0 &  4.0$\pm$0.7  &   12$\pm$6    \\
RCDS1252\_03 & 12:52:56.23 & $-$29:27:29.0 &  3.3$\pm$0.6  &   23$\pm$5    \\
RCDS1252\_04 & 12:52:58.99 & $-$29:27:01.0 &  2.6$\pm$0.6  &   19$\pm$6    \\
RCDS1252\_05 & 12:52:54.09 & $-$29:26:45.0 &  4.9$\pm$0.6  &   13$\pm$5    \\
RCDS1252\_06 & 12:53:02.35 & $-$29:26:13.0 &  4.0$\pm$0.8  &   16$\pm$7    \\
RCDS1252\_07 & 12:52:47.35 & $-$29:25:53.0 &  3.3$\pm$0.7  &  $<$5  \\
RCDS1252\_08 & 12:53:00.82 & $-$29:25:41.0 &  7.1$\pm$0.8  &   16$\pm$7    \\
RCDS1252\_09 & 12:52:49.49 & $-$29:25:41.0 &  4.9$\pm$0.7  &   15$\pm$6    \\
RCDS1252\_10 & 12:52:59.29 & $-$29:24:29.0 &  4.2$\pm$0.9  &  $<$5  \\
RXJ0849\_01 & 08:49:13.93 & $+$44:51:57.5 &  6.0$\pm$1.0  &   10$\pm$9    \\
RXJ0849\_02 & 08:48:58.50 & $+$44:52:25.6 &  7.5$\pm$0.8  &   15$\pm$6    \\
RXJ0849\_03 & 08:49:06.40 & $+$44:52:29.6 &  6.0$\pm$0.9  &   23$\pm$7    \\
RXJ0849\_04 & 08:49:07.53 & $+$44:53:49.6 &  6.0$\pm$1.0  &  $<$6  \\
RXJ0849\_05 & 08:49:01.51 & $+$44:54:25.6 &  4.0$\pm$1.0  &   13$\pm$8    \\
XMUJ2235\_01 & 22:35:27.42 & $-$25:57:26.0 &  5.0$\pm$1.0  &   13$\pm$8    \\
XMUJ2235\_02 & 22:35:21.49 & $-$25:56:58.0 &  5.2$\pm$0.9  &   31$\pm$7    \\
XMUJ2235\_03 & 22:35:15.56 & $-$25:57:02.0 &  6.7$\pm$0.9  &   19$\pm$7    \\
XMUJ2235\_04 & 22:35:30.09 & $-$25:56:30.0 &  5.0$\pm$1.0  &   13$\pm$9    \\
\hline
\end{tabular}
\label{table:S2}
\end{table}
 
\begin{table}
\setcounter{table}{1}
\caption{continued.}
\begin{tabular}{ lrrrr}
\hline
\multicolumn{1}{c}{ID} &
\multicolumn{1}{c}{R.A.} &
\multicolumn{1}{c}{Decl.} &
\multicolumn{1}{c}{$S_{850}$} &
\multicolumn{1}{c}{$S_{450}$} \\
\multicolumn{1}{c}{} &
\multicolumn{2}{c}{[J2000]} &
\multicolumn{1}{c}{[mJy]} &
\multicolumn{1}{c}{[mJy]} \\
\hline
\hline
XMUJ2235\_05 & 22:35:30.68 & $-$25:55:54.0 &  4.0$\pm$1.0  &   22$\pm$9    \\
XMUJ2235\_06 & 22:35:17.64 & $-$25:54:46.0 &  5.0$\pm$2.0  &   20$\pm$10   \\
XCSJ2215\_01 & 22:16:01.30 & $-$17:39:35.0 &  4.4$\pm$0.8  &   20$\pm$6    \\
XCSJ2215\_02 & 22:15:59.06 & $-$17:39:43.0 &  7.7$\pm$0.7  &   26$\pm$6    \\
XCSJ2215\_03 & 22:16:02.97 & $-$17:38:39.0 &  7.4$\pm$0.7  &   21$\pm$6    \\
XCSJ2215\_04 & 22:16:00.74 & $-$17:38:35.0 &  4.0$\pm$0.7  &    6$\pm$5    \\
XCSJ2215\_05 & 22:15:58.50 & $-$17:38:19.0 &  3.8$\pm$0.6  &   15$\pm$5    \\
XCSJ2215\_06 & 22:15:59.90 & $-$17:37:59.0 &  3.6$\pm$0.6  &   17$\pm$5    \\
XCSJ2215\_07 & 22:16:04.94 & $-$17:37:51.0 &  3.6$\pm$0.7  &   24$\pm$6    \\
XCSJ2215\_08 & 22:15:48.43 & $-$17:37:31.0 &  3.1$\pm$0.8  &  $<$5  \\
XCSJ2215\_09 & 22:15:59.90 & $-$17:37:19.0 &  4.0$\pm$0.7  &    6$\pm$5    \\
XCSJ2215\_10 & 22:16:06.89 & $-$17:36:27.0 &  4.3$\pm$0.9  &   16$\pm$7    \\
XDCPJ0044\_01 & 00:44:05.97 & $-$20:34:40.5 &  2.2$\pm$0.5  &    9$\pm$4    \\
XDCPJ0044\_02 & 00:43:57.99 & $-$20:34:32.5 &  4.2$\pm$0.6  &    6$\pm$5    \\
XDCPJ0044\_03 & 00:44:00.56 & $-$20:34:12.5 &  2.5$\pm$0.5  &   16$\pm$4    \\
XDCPJ0044\_04 & 00:44:10.24 & $-$20:34:16.5 &  2.8$\pm$0.6  &   12$\pm$5    \\
XDCPJ0044\_05 & 00:44:05.40 & $-$20:34:12.5 &  3.7$\pm$0.5  &   16$\pm$4    \\
XDCPJ0044\_06 & 00:44:03.69 & $-$20:33:48.5 &  4.2$\pm$0.5  &   12$\pm$4    \\
XDCPJ0044\_07 & 00:44:13.66 & $-$20:33:44.5 &  3.4$\pm$0.7  &    8$\pm$5    \\
XDCPJ0044\_08 & 00:44:05.97 & $-$20:33:16.5 &  6.0$\pm$0.5  &   22$\pm$4    \\
XDCPJ0044\_09 & 00:44:12.80 & $-$20:33:00.5 &  3.5$\pm$0.7  &  $<$4  \\
XDCPJ0044\_10 & 00:44:01.13 & $-$20:32:16.5 &  2.6$\pm$0.6  &  $<$4  \\
XDCPJ0044\_11 & 00:44:06.82 & $-$20:31:48.5 &  6.7$\pm$0.7  &   31$\pm$5    \\
XDCPJ0044\_12 & 00:44:09.96 & $-$20:31:20.5 &  4.5$\pm$0.9  &  $<$4  \\
\hline
\end{tabular}
\end{table}

\clearpage
\onecolumn
\begin{landscape}
\setlength\LTcapwidth{1.6\textwidth}
\begin{longtable}{lrrrrrrrrrrr}
\caption{Table of identifications and infrared properties of SCUBA-2-identified \submm\ sources with infrared counterparts. } \\
\hline
  \multicolumn{1}{c}{ID} &
  \multicolumn{1}{c}{IR RA} &
  \multicolumn{1}{c}{IR Decl.} &
  \multicolumn{3}{c}{Counterpart selection} &
  \multicolumn{1}{c}{$S_{3.6}$} &
  \multicolumn{1}{c}{$S_{4.5}$} &
  \multicolumn{1}{c}{$S_{5.8}$} &
  \multicolumn{1}{c}{$S_{8.0}$} &
  \multicolumn{1}{c}{$S_{24}$} &
  \multicolumn{1}{c}{$S_{\rm PACS}$\footnotemark[1]} \\
  \multicolumn{1}{c}{} &
  \multicolumn{2}{c}{[J2000]} &
  \multicolumn{1}{c}{MIPS} &
  \multicolumn{1}{c}{PACS} &
  \multicolumn{1}{c}{IRAC} &
  \multicolumn{1}{c}{[\uJy]} &
  \multicolumn{1}{c}{[\uJy]} &
  \multicolumn{1}{c}{[\uJy]} &
  \multicolumn{1}{c}{[\uJy]} &
  \multicolumn{1}{c}{[mJy]} &
  \multicolumn{1}{c}{[mJy]} \\
\hline
\hline
RXJ0152\_01a & 01:52:43.83 & $-$13:58:56.4 & 1 & 0 & 0 &  40$\pm$10  &  50$\pm$10  &  60$\pm$10  &  70$\pm$20  &  0.82$\pm$0.04  &  5.37$\pm$0.09  \\
RXJ0152\_01b & 01:52:44.12 & $-$13:58:52.3 & 1 & 0 & 1 &  24$\pm$8   &  30$\pm$9   &  50$\pm$10  &  40$\pm$10  &  0.49$\pm$0.03  & \ldots  \\
RXJ0152\_03a & 01:52:42.16 & $-$13:58:08.1 & 0 & 1 & 0 &  28$\pm$9   &  40$\pm$10  &  60$\pm$10  &  60$\pm$10  &  0.50$\pm$0.03  & 11.05$\pm$0.09  \\
RXJ0152\_03b & 01:52:42.04 & $-$13:58:02.7 & 1 & 1 & 0 &  23$\pm$8   &  22$\pm$8   &  12$\pm$7   &  11$\pm$7   &  0.44$\pm$0.03  &  6.36$\pm$0.09  \\
RXJ0152\_07a & 01:52:40.14 & $-$13:57:09.6 & 1 & 0 & 1 &  11$\pm$6   &  17$\pm$7   &  19$\pm$8   &  21$\pm$9   &  0.11$\pm$0.02  & \ldots  \\
RXJ0152\_10a & 01:52:32.98 & $-$13:57:07.4 & 0 & 1 & 0 &  19$\pm$7   &  22$\pm$8   &  21$\pm$8   &  22$\pm$9   &  0.34$\pm$0.03  &  5.12$\pm$0.09  \\
RXJ0152\_10b & 01:52:32.85 & $-$13:57:03.4 & 0 & 1 & 0 &  60$\pm$10  &  70$\pm$10  &  80$\pm$20  &  70$\pm$20  &  0.34$\pm$0.03  &  5.90$\pm$0.10  \\
RXJ0152\_13a & 01:52:41.11 & $-$13:55:56.2 & 1 & 0 & 1 &  40$\pm$10  &  60$\pm$10  &  70$\pm$10  &  60$\pm$10  &  0.55$\pm$0.03  & \ldots  \\
WARP1415\_01a & 14:15:10.82 & $+$36:11:00.1 & 0 & 0 & 1 &  50$\pm$10  &  50$\pm$10  &  40$\pm$10  &  50$\pm$10  & \ldots  & \ldots  \\
WARP1415\_07a & 14:15:13.47 & $+$36:12:10.9 & 0 & 0 & 1 & 300$\pm$30  & 260$\pm$30  & 170$\pm$20  & 180$\pm$20  & \ldots  & \ldots  \\
WARP1415\_10a & 14:15:11.42 & $+$36:13:05.1 & 0 & 0 & 1 &  31$\pm$9   &  40$\pm$10  &  40$\pm$10  &  30$\pm$10  & \ldots  & \ldots  \\
RCDS0910\_01a & 09:10:40.93 & $+$54:20:41.5 & 1 & 0 & 1 &  14$\pm$6   &  17$\pm$7   &  27$\pm$10  &   2$\pm$5   &  0.41$\pm$0.03  & \ldots  \\
RCDS0910\_01b & 09:10:40.69 & $+$54:20:45.0 & 1 & 0 & 0 &  22$\pm$8   &  27$\pm$9   &  19$\pm$8   &   3$\pm$5   &  0.50$\pm$0.04  & \ldots  \\
RCDS0910\_04a & 09:10:45.48 & $+$54:21:22.3 & 0 & 0 & 1 &   5$\pm$4   &   4$\pm$4   &  11$\pm$7   &  30$\pm$10  &  0.07$\pm$0.02  & \ldots  \\
RCDS0910\_11a & 09:10:54.86 & $+$54:22:18.1 & 1 & 0 & 0 &  50$\pm$10  &  60$\pm$10  &  80$\pm$20  & 110$\pm$20  &  1.32$\pm$0.05  & \ldots  \\
RCDS0910\_11b & 09:10:54.79 & $+$54:22:23.5 & 1 & 0 & 0 &  36$\pm$10  &  50$\pm$10  &  70$\pm$10  &  60$\pm$10  &  1.47$\pm$0.05  & \ldots  \\
RCDS1252\_02a & 12:52:47.83 & $-$29:27:52.9 & 0 & 0 & 1 &  40$\pm$10  &  50$\pm$10  &  50$\pm$10  &  40$\pm$10  & \ldots & \ldots  \\
RCDS1252\_08a & 12:53:00.99 & $-$29:25:40.7 & 0 & 0 & 1 & 160$\pm$20  & 120$\pm$20  &  90$\pm$20  &  60$\pm$10  & \ldots & \ldots  \\
RXJ0849\_02a\footnotemark[2] & 08:48:58.59 & $+$44:52:30.3 & 0 & 0 & 1 & 270$\pm$30  & 220$\pm$20  & 160$\pm$20  & 140$\pm$20  &  0.32$\pm$0.03  & \ldots  \\
RXJ0849\_04a & 08:49:07.62 & $+$44:53:50.1 & 1 & 0 & 1 &  36$\pm$10  &  60$\pm$10  & 110$\pm$20  & 220$\pm$30  &  0.56$\pm$0.04  & \ldots  \\
XMUJ2235\_02a & 22:35:21.48 & $-$25:56:58.5 & 0 & 0 & 1 & 110$\pm$20  & 130$\pm$20  & 120$\pm$20  &  80$\pm$20  & \ldots & \ldots  \\
XMUJ2235\_03a & 22:35:15.45 & $-$25:57:02.9 & 0 & 1 & 1 &  60$\pm$10  &  70$\pm$10  &  70$\pm$10  &  60$\pm$10  & \ldots &  4.26$\pm$0.07  \\
XCSJ2215\_01a & 22:16:01.19 & $-$17:39:35.4 & 0 & 1 & 1 &  80$\pm$10  &  70$\pm$10  & 120$\pm$20  & 160$\pm$20  &  0.80$\pm$0.10  &  5.03$\pm$0.06  \\
XCSJ2215\_02a & 22:15:59.01 & $-$17:39:42.6 & 1 & 1 & 1 &  40$\pm$10  &  60$\pm$10  & 110$\pm$20  & 150$\pm$20  &  0.67$\pm$0.10  &  9.43$\pm$0.06  \\
XCSJ2215\_03a & 22:16:02.73 & $-$17:38:39.2 & 0 & 1 & 0 &  \ldots   &  12$\pm$6   &   3$\pm$4   &   6$\pm$7   &  1.28$\pm$0.09  & 17.34$\pm$0.06  \\
XCSJ2215\_03b & 22:16:03.03 & $-$17:38:36.8 & 0 & 1 & 0 &  50$\pm$10  &  60$\pm$10  &  90$\pm$20  &  90$\pm$20  &  1.29$\pm$0.09  & 22.28$\pm$0.06  \\
XCSJ2215\_03c & 22:16:03.16 & $-$17:38:39.8 & 0 & 1 & 1 & 230$\pm$20  & 230$\pm$30  & 250$\pm$30  & 360$\pm$30  &  1.27$\pm$0.08  & 27.51$\pm$0.06  \\
XCSJ2215\_04a & 22:16:00.56 & $-$17:38:35.4 & 0 & 0 & 1 & 100$\pm$20  & 110$\pm$20  &  90$\pm$20  &  50$\pm$10  &  0.84$\pm$0.07  &  4.65$\pm$0.06  \\
XCSJ2215\_06a\footnotemark[3] & 22:15:59.71 & $-$17:37:59.0 & 0 & 0 & 1 & 130$\pm$20  & 170$\pm$20  & 120$\pm$20  & 100$\pm$20  &  0.51$\pm$0.07  &  4.46$\pm$0.06  \\
XCSJ2215\_07a & 22:16:04.75 & $-$17:37:51.9 & 0 & 1 & 0 &  33$\pm$9   &  40$\pm$10  &  40$\pm$10  &  30$\pm$10  &  0.25$\pm$0.08  &  8.91$\pm$0.07  \\
XCSJ2215\_07b & 22:16:04.97 & $-$17:37:54.3 & 0 & 1 & 0 & 100$\pm$20  &  70$\pm$10  &  60$\pm$10  &  60$\pm$10  &  0.28$\pm$0.09  &  8.23$\pm$0.07  \\
XCSJ2215\_09a & 22:15:59.99 & $-$17:37:18.2 & 1 & 0 & 0 &   8$\pm$5   &   8$\pm$5   &  10$\pm$6   &   4$\pm$5   &  2.10$\pm$0.20  & \ldots  \\
XDCPJ0044\_01a & 00:44:06.17 & $-$20:34:38.7 & 0 & 1 & 0 & 480$\pm$40  & 490$\pm$40  &   \ldots   &   \ldots   & \ldots & 45.19$\pm$0.09  \\
XDCPJ0044\_11a & 00:44:06.81 & $-$20:31:47.2 & 0 & 1 & 0 &  60$\pm$10  &  70$\pm$10  &   \ldots   &   \ldots   & \ldots &  7.80$\pm$0.10  \\
\hline
\footnotetext[1]{PACS data for RX\,J0152$-$1357 and RX\,J0849$+$4453 are at $70$\,\micron. The other fields are covered by $100$\,\micron\ data.}
\footnotetext[2]{Archival spectroscopic data for this source suggests a redshift of $z=1.589$, suggesting it is not a cluster member \citep{SDSSDR13}.}
\footnotetext[3]{Confirmed cluster member \citep{Stach2017}.}
\label{table:IR}
\end{longtable}
\end{landscape}
\clearpage
\twocolumn

\bsp	
\label{lastpage}
\end{document}